\documentclass[fleqn,usenatbib]{mnras}

\usepackage[T1]{fontenc}

\DeclareRobustCommand{\VAN}[3]{#2}
\let\VANthebibliography\thebibliography
\def\thebibliography{\DeclareRobustCommand{\VAN}[3]{##3}\VANthebibliography}

\usepackage{graphicx}	
\usepackage{amsmath}	
\usepackage{amssymb}	
\usepackage{multirow}
\usepackage{newtxtext,newtxmath}

\title[Hidden active nuclei unveiled by \ion{He}{ii}$\lambda$4686]{Unveiling hidden active nuclei in MaNGA star-forming galaxies with \ion{He}{ii}$\lambda$4686 line emission}

\author[G. Tozzi et al.]{
Giulia Tozzi$^{1,2}$\thanks{E-mail: giulia.tozzi@unifi.it (GT)},
Roberto Maiolino$^{3,4,5}$,
Giovanni Cresci$^{2}$,
Joanna M. Piotrowska$^{3,4}$,
Francesco Belfiore$^{2}$,
\newauthor
Mirko Curti$^{3,4}$,
Filippo Mannucci$^{2}$,
Alessandro Marconi$^{1,2}$
\\
$^{1}$Dipartimento di Fisica e Astronomia, Università di Firenze, Via G. Sansone 1, 50019, Sesto Fiorentino (Firenze), ITA\\
$^{2}$INAF - Osservatorio Astrofisico di Arcetri, Largo E. Fermi 5, 50127, Firenze, ITA\\
$^{3}$Kavli Institute for Cosmology, University of Cambridge, Madingley Road, Cambridge CB3 0HA, UK\\
$^{4}$Cavendish Laboratory, Astrophysics Group, University of Cambridge, 9 JJ Thomson Avenue, Cambridge CB3 0HE, UK\\
$^{5}$Department of Physics \& Astronomy, University College London, Gower Street, London WC1E 6BT, UK
}

\date{Accepted XXX. Received YYY; in original form ZZZ}

\pubyear{2023}

\begin{document}
\label{firstpage}
\pagerange{\pageref{firstpage}--\pageref{lastpage}}
\maketitle

\begin{abstract}
Nebular \ion{He}{ii}$\lambda$4686\AA~line emission is useful to unveil active galactic nuclei (AGN) residing in actively star-forming (SF) galaxies, typically missed by the standard BPT classification. Here we adopt the \ion{He}{ii} diagnostic to identify hidden AGN in the Local Universe using for the first time spatially-resolved data from the Data Release 15 of the Mapping Nearby Galaxies at APO survey (MaNGA DR15).
By combining results from \ion{He}{ii} and BPT diagnostics, we overall select 459 AGN host candidates ($\sim$10\% in MaNGA DR15), out of which 27 are identified as AGN by the \ion{He}{ii} diagram only. The \ion{He}{ii}-only AGN population is hosted by massive (M$_*\gtrsim10^{10}$ M$_{\odot}$) SF Main Sequence galaxies, and on average less luminous than the BPT-selected AGN. Given the \ion{He}{ii} line faintness, we revisit our census accounting for incompleteness effects due to the \ion{He}{ii} sensitivity limit of MaNGA. We thus obtain an overall increased fraction (11\%) of AGN in MaNGA compared to the BPT-only census (9\%), which further increases to 14\% for galaxies more massive than $10^{10}$ M$_{\odot}$; interestingly, on the SF Main Sequence the increase is by about a factor of 2. 
A substantial number of AGN in SF galaxies points to significant, coeval star formation and black hole accretion, consistently with results from hydrodynamical simulations and with important implications on quenching scenarios. In view of exploring unprecedented high redshifts with JWST and new ground-based facilities, revisiting the standard BPT classification through novel emission-line diagnostics is fundamental to discover AGN in highly SF environments.
\end{abstract}




\section{Introduction}

Active galactic nuclei (AGN) are considered crucial ingredients in baryonic structure formation in the Universe (e.g. \citealt{2003ApJ...599...38B, 2006MNRAS.365...11C, 2012ARA&A..50..455F}). Huge efforts have been invested in compiling large and complete AGN samples, including both high-luminosity quasars and low-luminosity Seyfert galaxies, to understand their physical properties and establish their role in galaxy evolution. Based on the distinctive characteristics of AGN (e.g. accretion rate and obscuration state), various techniques have been developed to search for AGN in different spectral bands, from hard X-rays (e.g. \citealt{2012ApJ...749...21A}) to mid-IR (e.g. \citealt{2005ApJ...631..163S, 2012ApJ...753...30S}) and radio (e.g. \citealt{2004A&A...416...35W}). The easiest systems to identify are the unobscured (type 1) AGN (e.g. \citealt{2010AJ....139.2360S}), while the selection of obscured (type 2) AGN is more challenging because the bright nuclear emission is not seen. In the optical, narrow emission-line flux ratios and diagnostic diagrams represent a widely used method to identify luminous type 2 AGN in optical spectra \citep{2003MNRAS.346.1055K, 2006MNRAS.372..961K}. Among these, a commonly used diagnostic is the BPT diagram \citep{1981PASP...93....5B}, based on the combined [\ion{O}{III}]/H$\beta$ and [\ion{N}{II}]/H$\alpha$ line ratios (with [\ion{S}{II}]/H$\alpha$ or [\ion{O}{I}]/H$\alpha$ as alternatives). 

Recently, an alternative diagnostics was presented by \cite{2012MNRAS.421.1043S}, the so-called \ion{He}{II} diagram, where the [\ion{O}{III}]$\lambda$5007\AA~ emission line flux is replaced by \ion{He}{II}$\lambda$4686\AA~ (hereafter [\ion{O}{III}] and \ion{He}{II}, respectively) in the ratio with the H$\beta$ line flux. Given the higher ionisation potential of \ion{He}{II} ($E_{\text{ion}}(He^{\text{+}})=54.4$ eV) compared to [\ion{O}{III}] ($E_{\text{ion}}(O^{\text{++}})=35.2$ eV), the \ion{He}{II} diagnostics is more sensitive to AGN activity and better differentiates between ionisation due to AGN and star formation. In particular, it allows to identify low-luminosity AGN in SF galaxies, where the AGN emission may be overwhelmed by intense star formation and, therefore, missed by the standard BPT-based selection, as pointed out by \citet{2010ApJ...711..284S}. \cite{2017MNRAS.466.2879B} indeed used the \ion{He}{II} diagram to search for nuclear activity in local galaxies in the Sloan Digital Sky Survey Data Release 7 (SDSS DR7; \citealt{2000AJ....120.1579Y, 2006AJ....131.2332G, 2009ApJS..182..543A}). 

Although the \ion{He}{ii} line emission is an optimal tracer of AGN activity, there are several limitations associated with its use. First of all, it is a faint emission line, which is the reason why the published studies on the topic are few and all limited to the Local Universe. Moreover, it can also be of stellar origin. The \ion{He}{II} line is indeed frequently observed in HII regions, associated with young stellar populations and, in particular, with evolved and massive Wolf-Rayet (WR) stars. However, in presence of WR stars, the \ion{He}{II} line appears blended with several metal lines, forming the so-called `blue-bump' around $\lambda$4650\AA~ \citep{2008A&A...485..657B}. Indeed,
several studies have also found evidence for substantial \ion{He}{ii} emission in low-metallicity, star forming dwarf galaxies \citep[e.g.][]{2022ApJ...930...37U}. In the case of dwarf galaxies some authors have also proposed other ionising sources, such as X-ray binaries \citep[e.g.][]{2019A&A...622L..10S} and fast radiative shocks \citep[e.g.]{1991ApJ...373..458G, 2005ApJS..161..240T}. However, recent observations have revealed the presence of AGN in several dwarf galaxies \citep[e.g.][]{2016ApJ...817...20M, 2020ApJ...898L..30M, 2022Natur.601..329S}, therefore indicating that AGN ionising photons may be responsible for the \ion{He}{ii} emission even in these systems. 

Most searches for AGN using optical spectroscopy have so far relied on single-fibre observations (such as the SDSS). As pointed out in \citet{2017MNRAS.467.2612W}, the small size of the optical fibres ($\sim3^{\prime\prime}$ diameter) can cover only the central region of the galaxy, where the AGN might be hidden due to obscuring material or dominated by other ionisation processes (like SF), and may miss extended visible Narrow Line Regions (NLR). In consequence, the central AGN may not be detected in the central integrated spectrum of its host, leading to an incorrect classification of the galaxy as non-active. Thanks to integral field unit (IFU) observations, it is now possible to search for weak (less diluted) and off-centre AGN signatures, such as cases of AGN offset from the centre due to a recent merger with a non-active galaxy, or of a second off-centre AGN in dual AGN systems \citep{2011ApJ...732....9G, 2012ApJ...753...42C, 2014ApJ...789..112C, 2016ApJ...823...42H, 2022NatAs.tmp..187M}. Another possibility might also be a recently turned off AGN \citep{2010A&A...509A.106S, 2016A&A...593L...8M}, whose relic ionisation remains visible at large distance from the centre \citep{2012MNRAS.420..878K, 2015AJ....149..155K}.

In this work, we combine the power of integral field spectroscopy in the MaNGA survey with the use the \ion{He}{ii} diagnostics to identify AGN host candidates. By comparing the AGN samples selected by the \ion{He}{ii} emission and the standard BPT diagrams, we find that the former is crucial to detect elusive AGN residing in highly SF galaxies, which are completely missed by the standard BPT classification, therefore increasing the total number of AGN in our parent sample.


This paper is organised as follows. In Sect. \ref{sec:selection}, we briefly introduce the MaNGA survey and describe how we identify AGN host candidates starting from the two different spatially-resolved diagnostics, that are the \ion{He}{ii} and BPT diagrams. In Sect. \ref{sec:host_properties}, we compare the main physical properties of the AGN host galaxies selected by the two diagnostics, while in Sect. \ref{sec:discussion} we discuss and quantify the impact of the \ion{He}{ii} detection limit in MaNGA on our overall AGN census. The conclusions from our study are finally drawn in Sect. \ref{sec:conclusion}.

Throughout this work we adopt a flat $\Lambda$CDM cosmology with $\Omega_{\text{m,0}} = 0.3$, $\Omega_{\Lambda,\text{0}} = 0.7$ and $H_0 = 70$ km s$^{-1}$ Mpc$^{-1}$.

\section{Sample selection and classification}
\label{sec:selection}
\subsection{MaNGA data}
\label{subsec:data}
Our sample is based on the fifteenth data release (DR15) of the MaNGA survey (Mapping Nearby Galaxies at APO; \citealt{2015ApJ...798....7B, 2015AJ....149...77D, 2015AJ....150...19L, 2016AJ....152..197Y, 2017AJ....154...86W}). MaNGA is an optical fibre-bundle IFU spectroscopic survey, part of the fourth phase of the SDSS (SDSS-IV; \citealt{2017AJ....154...28B}). The MaNGA DR15 catalogue consists of spatially resolved data of about 4600 local ($z\sim0.03$) galaxies with a spectral resolution $R\sim3000$ over the wavelength range of $\sim3600-10300$ \AA~, using multiple (from 19 to 127) fibre bundles. The median effective spatial resolution of MaNGA data is of $2.54^{\prime\prime}$ full width at half maximum \citep{2016AJ....152...83L}, corresponding to $\sim2$ kpc at $z\sim0.05$, and the pixel scale is of 0.5$^{\prime\prime}$.
In our analysis, we use spatially resolved data and measurements of galaxies (e.g. data and model cubes, emission-line flux maps) provided by the MaNGA data-analysis pipeline (DAP; \citealt{2019AJ....158..160B, 2019AJ....158..231W}), and take global properties (e.g. star formation rate, stellar mass) integrated within the field of view, from the MaNGA Pipe3D value added catalog \citep{2016RMxAA..52..171S}.





\subsection{Spatially resolved emission line diagrams}
\label{subsec:eldiag}

With the aim of comparing the efficacy of BPT and \ion{He}{ii} diagrams in selecting AGN galaxies, we start dealing with spatially-resolved measurements of emission-line flux of all MaNGA spaxels. From the MaNGA DAP we select spaxels with a signal-to-noise ratio (S/N) higher than 3 in [\ion{O}{iii}], [\ion{N}{ii}], H$\beta$ and H$\alpha$ for the BPT diagram (BPT spaxels), whereas we require S/N $>5$ in \ion{He}{ii} and S/N $>3$ in [\ion{N}{ii}], H$\beta$ and H$\alpha$ for the \ion{He}{ii} diagram (\ion{He}{ii} spaxels). 
We apply a higher S/N cut for the \ion{He}{ii}, compared to that for the other lines, in order to conservatively select only spaxels with a secure \ion{He}{ii} detection and to avoid cases of putative \ion{He}{ii} line emission that is actually consequence of a strong residual of the stellar continuum or to contamination from a foreground galaxy (these cases are discussed later in Sect. \ref{subsec:fake}.)

These S/N cuts lead to $\sim$2.1 million and $\sim$15200 spaxels for the BPT and \ion{He}{ii} diagrams, respectively (out of a total of more than 4.3 million MaNGA spaxels). The large discrepancy (i.e. $\sim$2 orders of magnitude) in the number of high-S/N spaxels resulting from the two separate S/N cuts (i.e. the BPT-selected versus \ion{He}{ii}--selected spaxels) reveals how faint and rare the detection of the \ion{He}{ii} emission line is, even in the spectra of local galaxies. 

To compute line ratios, we use dust-corrected emission line fluxes resulting from the Gaussian line modelling performed within the MaNGA DAP, except for a few galaxies which instead need a complete or partial spectral re-fitting. In the case of these objects, we take single-spaxel emission line fluxes obtained from our re-modelling (see Sect. \ref{subsec:refitting}).

\begin{figure*}
	\includegraphics[width=1.8\columnwidth]{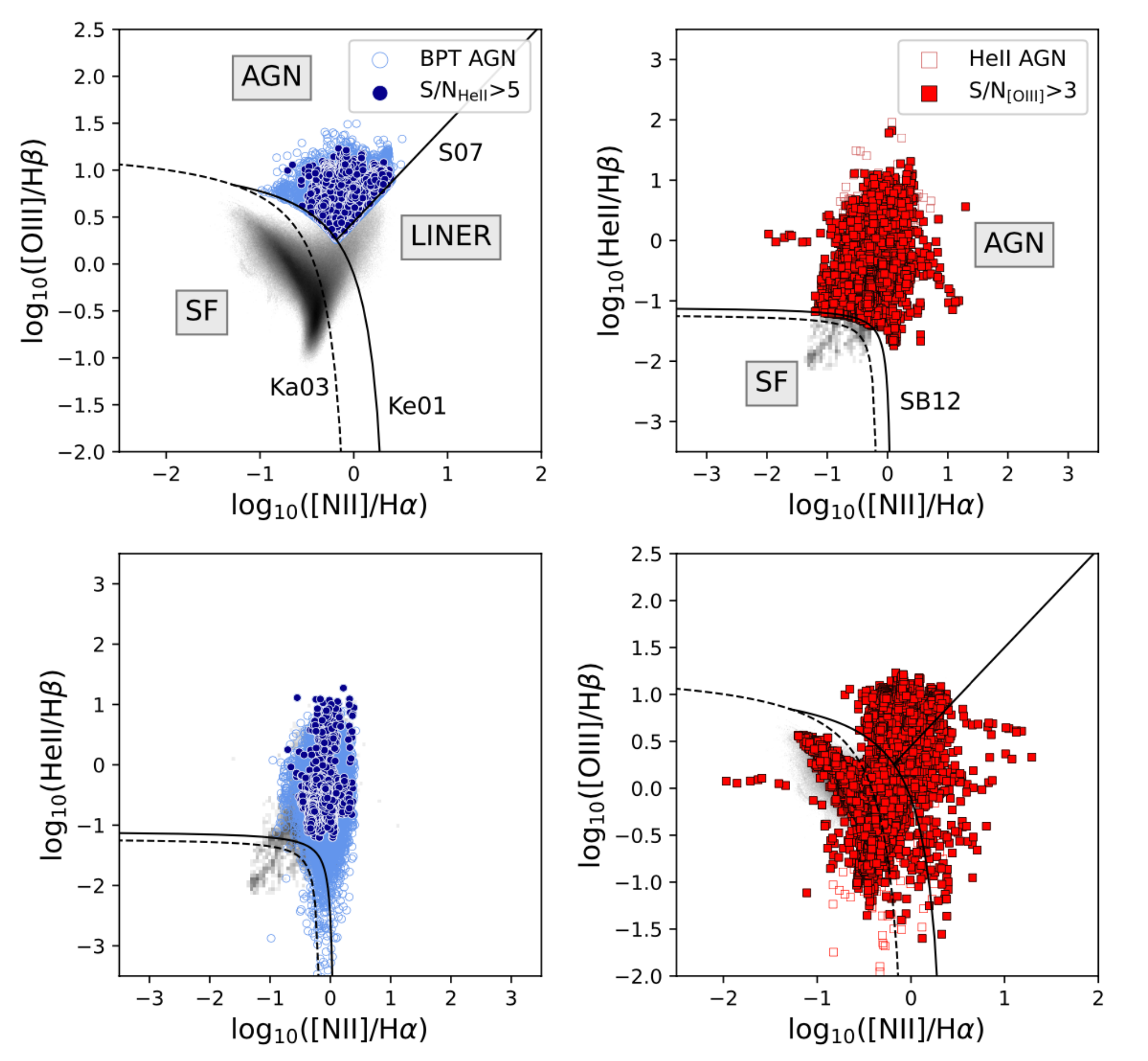}
    \caption{Spatially resolved emission line diagnostics diagrams. \textit{Top.} BPT (\textit{left}) and \ion{He}{ii} (\textit{right}) diagrams for MaNGA DR15 spaxels. The BPT AGN region is delimited by the Ke01 line, defined in \citet{2001ApJ...556..121K} as the extreme starburst line, and the S07 line separating AGN from LINER ionising sources \citep{2007MNRAS.382.1415S}. In the \ion{He}{ii} diagram the dashed and solid SB12 lines represent the limit above which 10\% and 50\% of the \ion{He}{ii} emission, respectively, is expected to come from an AGN \citep{2012MNRAS.421.1043S}. In each diagram the grey shading represents MaNGA spaxels satisfying the S/N threshold (Sect. \ref{subsec:eldiag}). Spaxels falling in the AGN region of the BPT and \ion{He}{ii} diagrams are shown as circles (BPT AGN) and squares (\ion{He}{ii} AGN), respectively. Filled symbols represent the two categories meeting the additional requirement of having S/N(\ion{He}{ii}) $>5$ ($\sim$3440 spaxels) and S/N([\ion{O}{iii}]) $>3$ ($\sim$11700 spaxels), respectively. \textit{Bottom.} Distribution of BPT AGN and \ion{He}{ii} AGN spaxels in the \ion{He}{ii} (\textit{left}) and BPT diagram (\textit{right}), respectively. BPT AGN are mostly all confirmed by the \ion{He}{ii} diagram, while \ion{He}{ii}-selected AGN spread largely also in the SF region of the BPT diagram.}
    \label{fig1}
\end{figure*}

\subsubsection{BPT and \ion{He}{ii} diagnostic diagrams}
\label{subsec:bpt_and_heii}
The top panels of Fig. \ref{fig1} show the [\ion{O}{iii}]/H$\beta$ versus [\ion{N}{ii}]/H$\alpha$ version of the BPT (left panel) and the \ion{He}{ii} (right) diagnostic diagrams of MaNGA spaxels. 

Regions in the two diagrams corresponding to different ionisation mechanisms are delimited by demarcation lines. In the BPT diagram, the Ke01 and Ka03 lines represent the theoretical extreme starburst line \citep{2001ApJ...556..121K} and the empirical SF line \citep{2003MNRAS.346.1055K}, respectively; the S07 line instead separates LINERs from AGN ionising sources according to \citet{2007MNRAS.382.1415S}. Similarly, we plot separating curves in the \ion{He}{ii} diagram as determined in \citet{2012MNRAS.421.1043S}: the solid and dashed lines are defined as the limiting curves above which more than 50\% and 10\% of the \ion{He}{ii} flux, respectively, is expected to come from an AGN. We classify spaxels as AGN-like only when they lie in the pure AGN regions of their respective diagrams: in the BPT, these are the spaxels falling above both the Ke01 and S07 lines, referred to as BPT AGN spaxels (empty/filled blue circles); while in the \ion{He}{ii} diagram, the \ion{He}{ii} AGN spaxels are those above the 50\% AGN line (empty/filled red squares). Filled symbols represent BPT and \ion{He}{ii} AGN spaxels meeting the additional requirement of having S/N(\ion{He}{ii}) $>5$ and S/N([\ion{O}{iii}]) $>3$, respectively. Whereas there is a substantial difference in number between filled ($\sim$3440 spaxels) and empty ($\sim$37700 spaxels) blue circles, almost all \ion{He}{ii} spaxels fulfilling the S/N threshold in \ion{He}{ii} ($\sim$12000 spaxels) have also S/N $>3$ in [\ion{O}{iii}] ($\sim$11700 spaxels). The grey shading in the background represents the total sample of MaNGA spaxels as selected by the S/N cut, namely the total BPT (left) and \ion{He}{ii} (right) spaxels as previously defined.

\subsubsection{Comparing the two diagnostic diagrams}
\label{subsec:bpt_vs_heii}

Similarly to the analysis performed on integrated emission line flux ratios in \citet{2017MNRAS.466.2879B}, in the bottom panels of Fig. \ref{fig1} we compare the spatial distributions of AGN spaxels identified with one diagnostic diagram, within the parameter space of the other diagnostic. More precisely, we show the distribution of BPT-selected AGN spaxels in the \ion{He}{ii} diagram (empty/filled blue circles, bottom left panel) and the distribution of \ion{He}{ii}-selected AGN spaxels in the BPT diagram (empty/filled red squares, bottom right panel).
Except for a very few cases, almost all BPT AGN spaxels fall above the 10\% AGN demarcation line of the \ion{He}{ii}-diagram, and 82\% of the total BPT AGN spaxels is located even above the 50\% AGN line in the \ion{He}{ii} diagram. Such a fraction increases to 100\%, if we consider only BPT AGN spaxels with also S/N(\ion{He}{ii}) $>5$ (filled blue circles). Therefore, the \ion{He}{ii}-classification overall recovers the BTP-classified AGN. 
Unlike the BPT AGN spaxels in the \ion{He}{ii} diagram, the \ion{He}{ii} AGN spaxels are scattered across the BPT plane and extend well into the SF region, with only 29\% of them meeting classification criteria for BPT AGN. Hence, Fig. \ref{fig1} clearly demonstrates the power of \ion{He}{ii} diagnostic in identifying regions with significant contribution to ionisation from an AGN, which may be erroneously classified as entirely due to star formation.

\subsection{Selecting AGN host galaxies in MaNGA}
\label{subsec:AGN_galaxies}

We define a criterion to identify AGN host galaxies based on their content of AGN spaxels as classified by the emission line diagnostic diagrams in Fig. \ref{fig1}. We identify as BPT (\ion{He}{ii}) AGN galaxy candidates those objects with at least 20 AGN spaxels meeting the S/N threshold, defined for the BPT (\ion{He}{ii}) diagram in Sect. \ref{subsec:eldiag}. The minimum requirement of 20 AGN spaxels approximately corresponds to one spatial resolution element in MaNGA, in case of adjacent spaxels. We check contiguity among spaxels at a later stage of the analysis (see Sect. \ref{subsec:fake} and Appendix \ref{appx1}).

With this definition we obtain populations of BPT and \ion{He}{ii} AGN galaxy candidates based on the BPT and \ion{He}{ii} spaxels classification, respectively. Among the \ion{He}{ii} AGN population, some galaxies are also BPT-selected AGN (hereafter BPT\&\ion{He}{ii} AGN); while others are identified as AGN only by the \ion{He}{ii} diagram (hereafter \ion{He}{ii}-only AGN), with the majority of their spaxels located in BPT SF region.

Details on the demography of the different AGN populations are provided at the end of Sect. \ref{subsec:type1_WR}, after refining our selected sample of AGN galaxies.

\subsubsection{Excluding cases of dubious \ion{He}{ii} line emission}
\label{subsec:fake}
As already mentioned, the main problem related to the use of the \ion{He}{ii} emission line is its faintness, limiting its clear detection to only a small sample of MaNGA galaxies.
From our preliminary selection we obtain 178 \ion{He}{ii} AGN galaxy candidates, most of which must be rejected because the \ion{He}{ii} emission line at S/N$>$5 (suspiciously reported by the MaNGA DAP even in spaxels in the galaxy outskirts) does not appear to be real. By visually inspecting single-spaxel spectra using Marvin\footnote{Online, interactive tool to search for, access and visualise MaNGA data (https://dr15.sdss.org/marvin/).} \citep{2019AJ....158...74C} and \ion{He}{ii} DAP maps, we indeed find, in $\sim$50\% of the selected galaxies, spaxels with a putative high-S/N \ion{He}{ii}, where the \ion{He}{ii} emission line is actually resulting from either noise, artifacts or strong residuals after the stellar continuum subtraction.

To identify and exclude these galaxies from our \ion{He}{ii} AGN sample, for each galaxy we plot the \ion{He}{ii} flux map and create the `subtracted cube', that is the MaNGA datacube after subtracting the DAP stellar continuum model. From the subtracted cube and the DAP emission line model cube\footnote{Both the stellar continuum and emission line models are stored in the MaNGA model cube produced by the MaNGA DAP.} of each galaxy, we then extract the spectrum and the cumulative emission line model using two different central apertures (i.e. $2.5^{\prime\prime}$ and $5.5^{\prime\prime}$). By visually examining both the \ion{He}{ii} flux maps and the integrated spectra, we finally exclude cases of \ion{He}{ii} non-detection from our preliminary selected sample.

We additionally report on five cases of contamination by a foreground galaxy (8084-12701, 8158-1901, 8158-1902, 8987-6101, 9194-6104), producing an emission line (likely H$\beta$) around the \ion{He}{ii} wavelength in the rest-frame of the target galaxy and, therefore, misinterpreted as the \ion{He}{ii} line by the DAP. These five galaxies are therefore excluded.

\subsubsection{Need for re-modelling some MaNGA datacubes}
\label{subsec:refitting}

\begin{figure}
	\includegraphics[width=0.95\columnwidth]{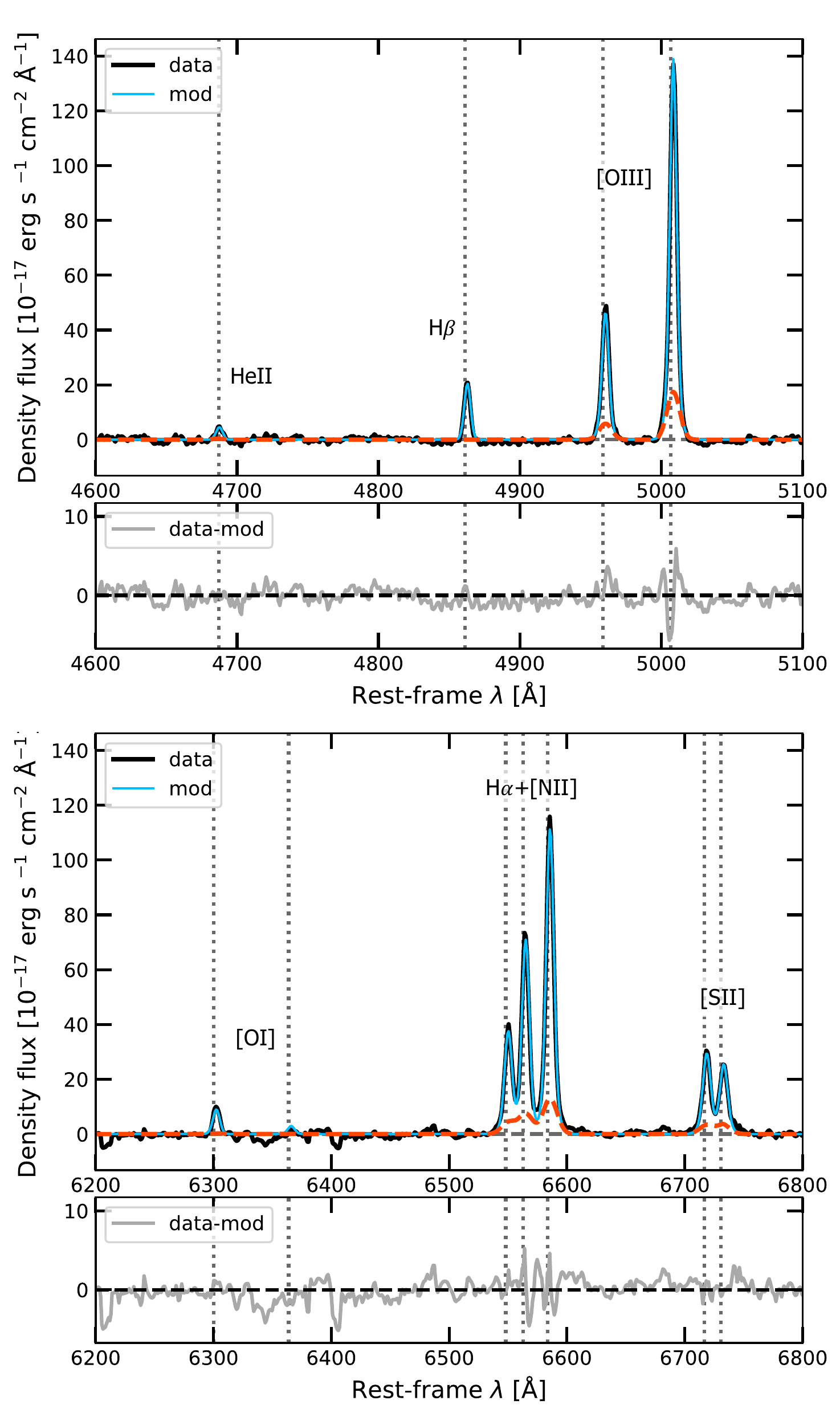}
    \caption{Best-fit results from our spectral re-modelling of the main emission lines in the MaNGA galaxy 9026-9101: \ion{He}{ii}, H$\beta$ and [\ion{O}{iii}] emission lines in the upper panel, whereas [\ion{O}{i}], [\ion{N}{ii}], H$\alpha$ and [\ion{S}{ii}] in the lower panel. In each panel, the subtracted data (black) and the total emission-line model (lightblue) are extracted from a central 2.5$^{\prime\prime}\times$2.5$^{\prime\prime}$ aperture. Two Gaussian components are required to accurately model the narrow-line emission in this galaxy, one of which is separately shown in orange. Below each main panel, we show corresponding residuals.}
    \label{fignew}
\end{figure}

\begin{table}
\centering
\setlength{\tabcolsep}{6pt} 
\renewcommand{\arraystretch}{1} 
\caption{List of refitted MaNGA datacubes. For each refitted datacube, the columns are arranged in the following order: (1) plate-ifu ID of the corresponding galaxy, (2) type of remodelling, (3) \ion{He}{ii}-based classification (AGN or SF) after our remodelling, and (4) identification of \ion{He}{ii}-only AGN. In column (2) we distinguish among full refitting of MaNGA datacubes (full), modelling of all emission lines in MaNGA continuum-subtracted datacubes (all lines), and re-modelling of the \ion{He}{ii} emission line only (\ion{He}{ii}).}
\begin{tabular}{cccc}
\hline
    Plate-ifu & Refitting & \ion{He}{ii} & \ion{He}{ii}-only\\
    ID & type & classification & AGN\\
 \hline
 7815-6104 & full & AGN & no\\
 8146-1901 & \ion{He}{II} & AGN & yes\\
 8320-9101 & all lines & SF & -\\
 8341-12704 & full & AGN & no\\
 8341-12705 & \ion{He}{II} & SF & -\\
 8458-3702 & \ion{He}{II} & SF & -\\
 8465-3701 & \ion{He}{II} & AGN & yes\\
 8615-3701 & full & AGN & yes\\
 8715-3702 & full & AGN & no\\
 8934-3701 & \ion{He}{II} & SF & -\\
 8940-6102 & \ion{He}{II} & AGN & yes\\
 9026-3701 & \ion{He}{II} & SF & -\\
 9026-9101 & all lines & AGN & no\\
 9487-3702 & full & AGN & no\\
 9487-9102 & full & AGN & yes\\
 9883-3701 & \ion{He}{II} & SF & -\\ 
   \hline
\end{tabular}
\label{tab1}
\end{table}

By visually inspecting extracted spectra along with their cumulative emission line model (as previously explained in Sect. \ref{subsec:fake}), we also find 16 MaNGA datacubes of \ion{He}{ii} AGN galaxies badly fitted by the MaNGA DAP (Table \ref{tab1}). We therefore remodelled all them by means of the fitting code for IFU data described in \citet{2020A&A...644A..15M} (see also \citealt{2021A&A...648A..99T}, \citealt{2023arXiv230111060C} for more details).

The fitting procedure by \citet{2020A&A...644A..15M} first models the full datacubes with \textsc{pPXF} \citep{2004PASP..116..138C,2017MNRAS.466..798C}, creating dedicated templates for each contributing spectral component: namely, stellar continuum, AGN continuum, and line emission from AGN broad and narrow line regions (BLR and NLR, respectively). The stellar continuum is modelled with the MILES extended stellar population templates \citep{2016A&A...589A..73R}, and the AGN continuum through a nth-degree polynomial. Finally, multiple Gaussian components are used to reproduce emission lines, including a broad component to account for any possible BLR contribution to permitted emission lines (such as H$\alpha$, H$\beta$ and \ion{He}{ii}). Each set of (broad or narrow) Gaussian components is constrained to have the same kinematics (i.e. same velocity $v$ and velocity dispersion $\sigma$). Then, the fitting code allows the user to subtract from the data the total continuum (i.e. stellar plus AGN continuum) and the unresolved
BLR line emission, thus creating a datacube containing only spatially-resolved narrow emission lines. At this point, a refined multiple Gaussian fitting of the narrow emission lines can be performed, with no longer contamination from continuum and/or unresolved emission.

Among the 16 badly fitted MaNGA datacubes, six galaxies are likely Seyfert 1.8 since their hydrogen Balmer lines clearly exhibit a very broad component originating in the AGN BLR, which the MaNGA DAP is not designed to reproduce. We therefore fully re-fitted these six datacubes (labelled as `full' in Table \ref{tab1}), starting from the modelling of the stellar continuum and including also a very broad Gaussian component ($\sigma>1000$ km s$^{-1}$) in the model of the permitted lines, to account for the BLR contribution to total line emission. In the remaining ten datacubes instead, no BLR line emission is present and the stellar continuum has been correctly modelled by the MaNGA DAP. We hence created the corresponding subtracted datacubes (as defined in Sect. \ref{subsec:fake}) and re-modelled emission lines only. In eight datacubes, the inaccurate best-fit model is limited to the faint \ion{He}{ii} emission line, while the other brighter emission lines have been correctly modelled. So, in these cases we re-fitted the \ion{He}{ii} emission line only (labelled as `\ion{He}{ii}' in Table \ref{tab1}), to obtain a more reliable flux. Two galaxies instead require a re-modelling of their entire subtracted cube, hence we re-fitted all emission lines (`all lines' in Table \ref{tab1}). In Fig. \ref{fignew}, we plot as representative example the best-fit results of the main emission lines in the MaNGA galaxy 9026-9101, as obtained from our spectral re-modelling: the subtracted data (black) and the total emission-line model (lightblue) have been extracted from a central 2.5$^{\prime\prime}\times$2.5$^{\prime\prime}$ aperture. To accurately model the narrow-line emission in this galaxy, we used two Gaussian components, one of which is separately shown (orange) in Fig. \ref{fignew}.

After the re-modelling, we create flux maps of the main emission lines and correct line fluxes for dust extinction \citep{2000ApJ...533..682C}. At this point, six galaxies (8320-9101, 8341-12705, 8458-3702, 8934-3701, 9026-3701, 9883-3701) are no longer classified as AGN galaxies (but as SF galaxies) according to the \ion{He}{ii} diagram, so we exclude them. In Table \ref{tab1}, we list the refitted MaNGA datacubes along with the type of remodelling. For each galaxy, we also indicate whether the \ion{He}{ii} AGN classification is confirmed after the remodelling, and we also identify \ion{He}{ii}-only AGN galaxies among such confirmed cases.

\subsubsection{Search for type 1 AGN and \ion{He}{ii} due to WR stars}
\label{subsec:type1_WR}

\begin{figure}
	\includegraphics[width=0.95\columnwidth]{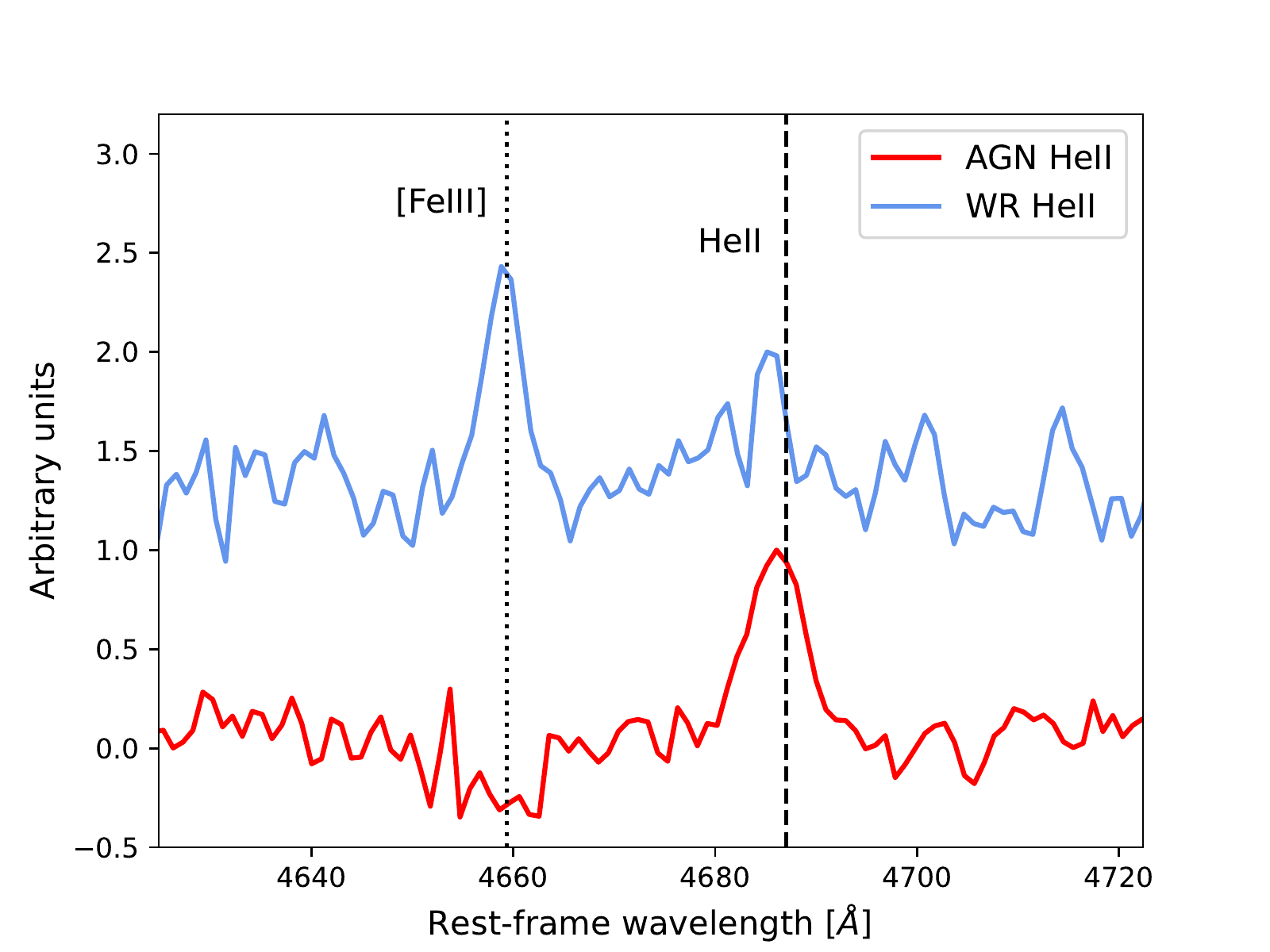}
    \caption{Representative examples of \ion{He}{ii} emission line associated with WR stars (blue) and with an AGN (red), respectively. The former is extracted from a WR region in the galaxy 8458-3702 (previously classified as a WR galaxy by \citealt{2020ApJ...896..121L}) and shows typical signatures of WR emission (i.e. blue bump around \ion{He}{ii} and bright [\ion{Fe}{iii}]$\lambda$4658\AA); the latter from the nuclear region of the galaxy 8257-12701, identified as \ion{He}{ii}-only AGN, clearly exhibiting a peaked \ion{He}{ii} emission line, AGN classified by the \ion{He}{ii} diagram.}
    \label{fig2}
\end{figure}

We finally look for type 1 AGN in the overall (BPT and/or \ion{He}{ii}) AGN sample, and for central \ion{He}{ii} emission due to WR stars among \ion{He}{ii} AGN galaxies. To identify both categories we compare the average velocity dispersion $\left<\sigma\right>$ of different emission lines within a 5.5$^{\prime\prime}$ central aperture. In particular, we classify as type 1 AGN those with $\left<\sigma(\text{H}\alpha)\right>>2\left<\sigma([\ion{O}{iii}])\right>$ or $\left<\sigma(\text{H}\beta)\right>>2\left<\sigma([\ion{O}{iii}])\right>$, resulting in 13 galaxies. We decide to exclude the type 1 AGN since we are primarily interested in AGN likely missed by the BPT classification, while these are very clear, luminous AGN. Furthermore, their datacubes should be refitted (just like the Seyfert 1.8 described in Sect. \ref{subsec:refitting}) and the determination of their SFR and stellar masses would be more uncertain.

Regarding \ion{He}{ii} emission due to WR stars, we find clear evidence for blue blump in three galaxies \ion{He}{ii}-classified as SF after our remodelling (8458-3702, 8934-3701, 9883-3701), identified as WR galaxies by \citet{2020ApJ...896..121L}. In Fig. \ref{fig2}, we show for comparison two spectra over the \ion{He}{ii} wavelength range, respectively extracted from a WR region in the galaxy 8458-3702 (blue), and from the nuclear region of the galaxy 8257-12701 (red), classified as \ion{He}{ii}-only AGN in our census. Whereas the former spectrum clearly exhibits a blue bump around the \ion{He}{ii} line along with a bright [\ion{Fe}{iii}]$\lambda$4658\AA~ typically associated with WR stars (e.g. \citealt{2021ApJ...915...21R}), the latter distinctly shows a peaked \ion{He}{ii} emission line, classified as AGN-like by the \ion{He}{ii} diagram.

However, we point out that this study does not aim at a rigorous identification of WR regions within MaNGA galaxies but, mostly, at excluding cases of clear WR \ion{He}{ii} emission originating in the galaxy centre, but misclassified as AGN-like based on the \ion{He}{ii} diagnostic. For this reason, we search in the remaining \ion{He}{ii} AGN sample for signatures of blue bump only within a central 5.5$^{\prime\prime}$ aperture, by requiring $\left<\sigma(\ion{He}{ii})\right>>2\left<\sigma([\ion{O}{iii}])\right>$ and $\left<\sigma(\ion{He}{ii})\right>>500$ km s$^{-1}$ (indicative lower limit taken from \citealt{2008A&A...485..657B}). We thus tentatively detect one case of blue bump in the central region of the galaxy 8250-6101. Yet, given the marginal detection of such bump compared to the narrow, nebular \ion{He}{ii} emission line, we decide to retain this in our AGN sample.

\begin{figure}
	\includegraphics[width=0.95\columnwidth]{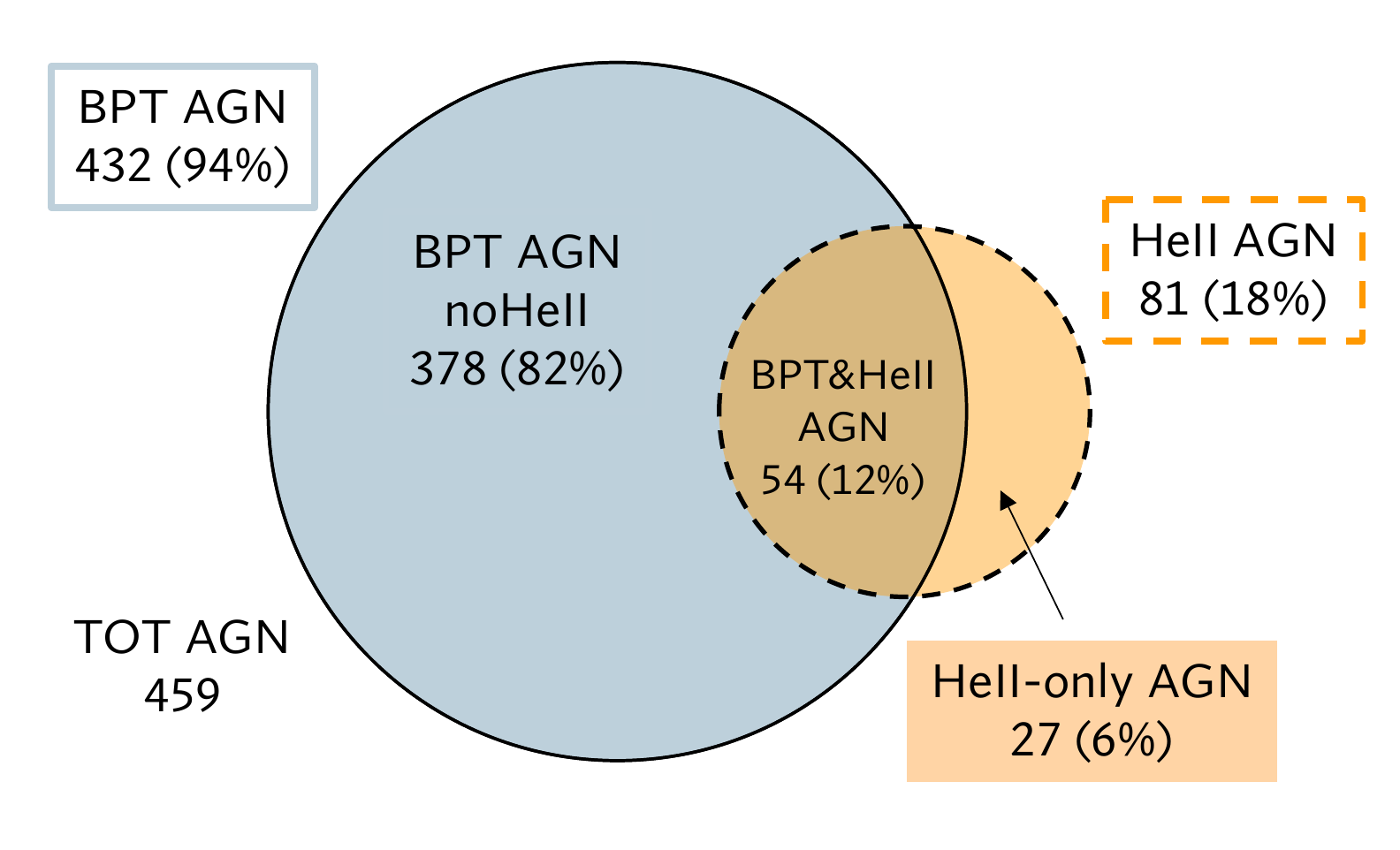}
    \caption{Chart summarising the demography of the AGN population selected in MaNGA. In total, we identify 459 AGN galaxy candidates by combining the two diagnostics: 432 (94\% out of our total AGN population) are BPT-classified AGN, while the \ion{He}{ii} diagram selects 81 (18\%) galaxies as AGN. 54 galaxies are selected by both diagrams (BPT\&\ion{He}{ii} AGN), corresponding to 12\% of the total AGN population, and 27 galaxies (6\% of the AGN population) are instead classified as AGN only by the \ion{He}{ii} diagnostic (\ion{He}{ii}-only AGN). Finally, there are 378 BPT AGN (82\%), with no \ion{He}{ii} detection (BPT AGN no\ion{He}{ii}).}
    \label{fig3}
\end{figure}

Figure \ref{fig3} summarises the demography of the AGN population (and subpopulations), consisting of 459 AGN host candidates in total ($\sim$10\% in MaNGA DR15), as resulting from our selection criterion. Out of the total AGN galaxies, 432 are BPT-selected AGN (94\%), whereas 81 are \ion{He}{ii}-selected AGN (18\%).
In the \ion{He}{ii} AGN subsample, 54 objects are classified as AGN galaxies also by the BPT (12\% and 67\% of the total and \ion{He}{ii} AGN samples, respectively). There are instead 27 \ion{He}{ii}-only AGN candidates (6\% and 33\% of the total and \ion{He}{ii} AGN samples, respectively), missed by the BPT AGN classification. There are finally 378 BPT AGN (82\%), with no \ion{He}{ii} detection (BPT AGN no\ion{He}{ii}). In Appendix \ref{appx1}, we show for every selected \ion{He}{ii}-only AGN galaxy spatially resolved maps displaying the spaxel classification according to BPT (Fig. \ref{fig_a1}) and \ion{He}{ii} (Fig. \ref{fig_a2}) classification, respectively, supporting the presence a central AGN and confirming their identification as \ion{He}{ii}-only AGN. By crossmatching our sample of \ion{He}{ii}-only AGN with the AGN catalogue reported by \citet{2020ApJ...901..159C}, we find that 11 out of 27 (41\%) \ion{He}{ii}-only AGN are identified as AGN based on their radio observations (10 objects) and/or mid-IR colours (2), further supporting the use of \ion{He}{ii} to select AGN missed in usual BPT diagnostics. Moreover, \citet{2020ApJ...901..159C} classify as AGN also the galaxy 8615-3701, due to the presence of broad emission lines in its SDSS spectrum. This object is indeed one of the MaNGA galaxies we identified as intermediate-type AGN and, for this, fully refitted in Sect. \ref{subsec:refitting}.

\subsection{Verifying the AGN-like nature of \ion{He}{ii} line emission}
\label{subsec:offset_AGN}

\begin{figure}
	\includegraphics[width=0.95\columnwidth]{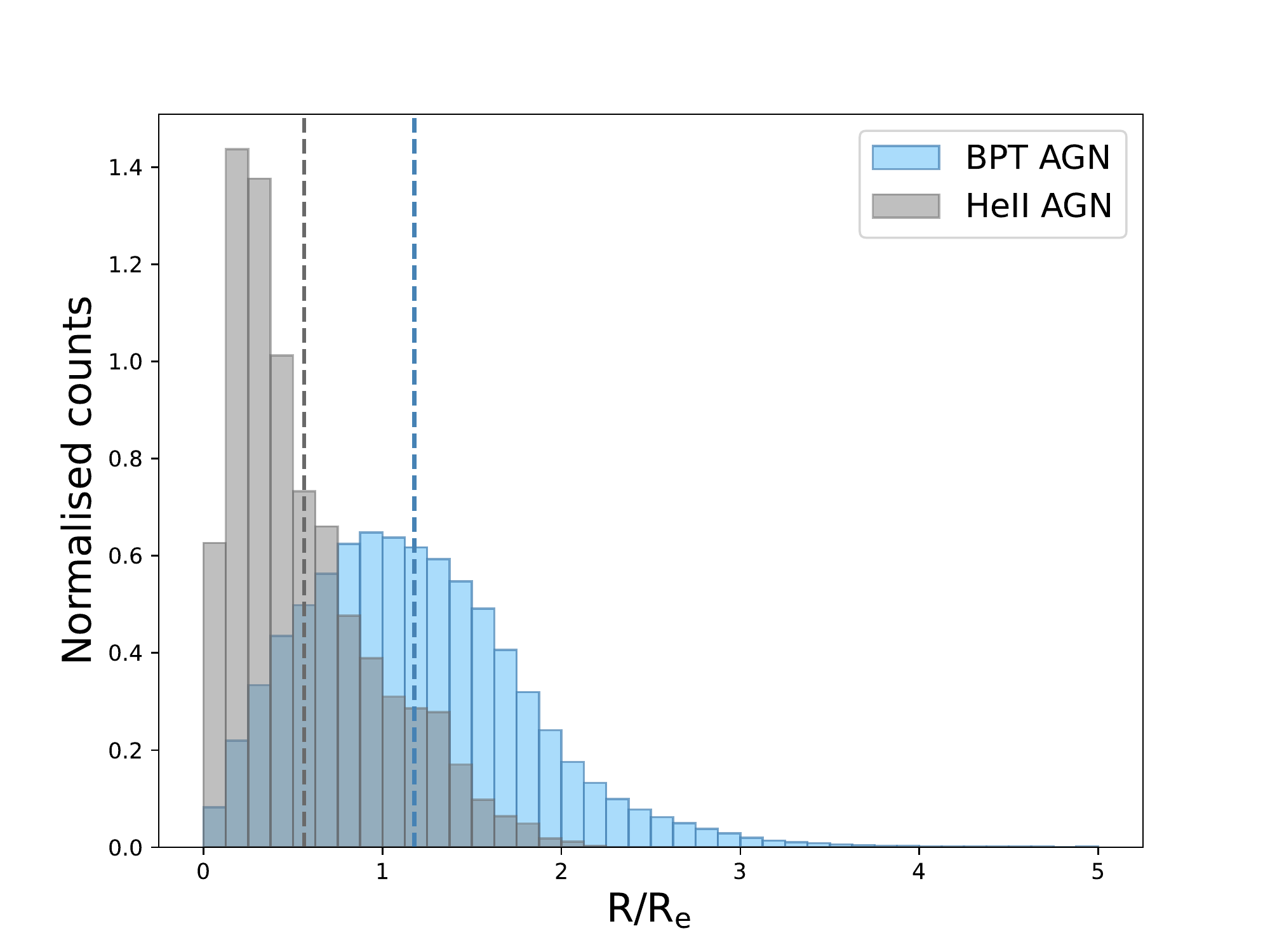}
    \caption{Normalised distributions of radial distance (in units of effective radius, R$_\mathrm{e}$) from the galaxy centre for the spaxels separately classified as BPT AGN (lightblue) and \ion{He}{ii} AGN (grey). The \ion{He}{ii} AGN spaxels are concentrated in the centre, with a mean radial distance of $\sim0.56$ R$_\mathrm{e}$ (dashed grey line), compared to the BPT-classified AGN spaxels ($\sim1.2$ R$_\mathrm{e}$, dashed lightblue line).}
    \label{fig4}
\end{figure}

\begin{figure*}
	\includegraphics[width=1.95\columnwidth]{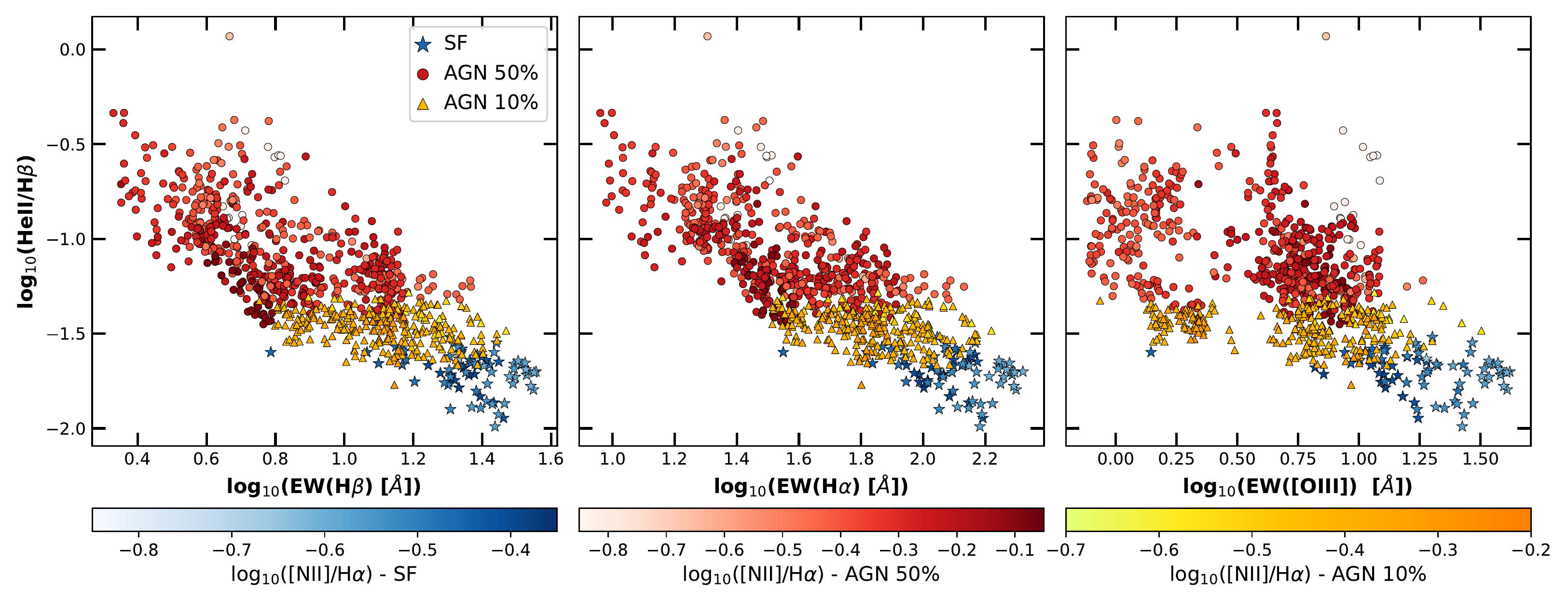}
    \caption{\ion{He}{ii}/H$\beta$ ratio as a function of EW(H$\beta$) (left panel), EW(H$\alpha$) (middle panel) and EW([\ion{O}{iii}]) (right panel) for all spaxels of \ion{He}{ii}-only AGN (plotted in Fig. \ref{fig_a2}), fulfilling the S/N threshold defined in Sect. \ref{subsec:eldiag}. We also show the dependence on the [\ion{N}{ii}]/H$\alpha$ ratio using different colourbars according to their \ion{He}{ii}-based classification as SF (star symbols, blue colourbar), strong AGN (squares, red colourbar), or weak AGN (triangles, yellow-to-orange colourbar). The \ion{He}{ii}/H$\beta$ ratios overall decrease with line EWs, with the largest EWs found in SF spaxels, consistently with EW(H$\beta$) values inferred for SDSS SF galaxies showing WR signatures \citep{2012MNRAS.421.1043S}. Such low-\ion{He}{ii}/H$\beta$ and high-EW(H$\beta$) SF spaxels also exhibit the lowest [\ion{N}{ii}]/H$\alpha$ ratios in agreement with those typical of SF galaxies \citep{2011MNRAS.413.1687C}.}
    \label{fig5}
\end{figure*}

We take advantage of the spatially resolved information of MaNGA to verify the AGN-like nature of the detected \ion{He}{ii} line emission. In Fig. \ref{fig4}, we plot the radial distribution (distance from the centre in units of effective radius, R$_\mathrm{e}$) of the AGN-like BPT (lightblue) and \ion{He}{ii} (grey) spaxels. In agreement with the AGN interpretation, the AGN-like \ion{He}{ii} emission mainly originates in the galaxy center.
In fact, the BPT AGN spaxels are on average at a distance of $\sim1.2$ R$_\mathrm{e}$ (dashed lightblue line) and located up to $\sim3$ R$_\mathrm{e}$ from the galaxy centre; while the \ion{He}{ii} AGN spaxels are exclusively found at shorter distances, where the S/N is high enough to detect \ion{He}{ii}, with a mean (dashed grey line) and maximum radial distance of $\sim0.56$ R$_\mathrm{e}$ and $\sim2$ R$_\mathrm{e}$, respectively.

As a final check on the correct selection of AGN-like spaxels, hence on the overall identification of \ion{He}{ii}-only AGN galaxies, in Fig. \ref{fig5} we plot \ion{He}{ii}/H$\beta$ ratio as a function of equivalent width (EW) of H$\beta$ (left panel), H$\alpha$ (middle panel) and [\ion{O}{iii}] (right panel) lines, for all spaxels of \ion{He}{ii}-only AGN fulfilling the S/N threshold (they are plotted in Fig. \ref{fig_a2}), defined in Sect. \ref{subsec:eldiag}. In addition, we show the dependence on the [\ion{N}{ii}]/H$\alpha$ ratio using different colourbars according to their \ion{He}{ii}-based classification as SF (star symbols, blue colourbar), strong AGN (squares, red colourbar), or weak AGN (triangles, yellow-to-orange colourbar). The \ion{He}{ii}/H$\beta$ ratios overall decrease at increasing line EWs, with the largest EWs associated SF spaxels, consistently with EW(H$\beta$) values inferred for SDSS SF galaxies showing WR signatures \citep{2012MNRAS.421.1043S}. Such SF spaxels, featured by low-\ion{He}{ii}/H$\beta$ and high-EW values, also exhibit the lowest [\ion{N}{ii}]/H$\alpha$ ratios in agreement with those typical of SF galaxies \citep{2011MNRAS.413.1687C}. In addition to have higher \ion{He}{ii}/H$\beta$ ratios (as already pointed out by the \ion{He}{ii} diagram), AGN spaxels are instead overall characterised by larger EW(H$\alpha$) ($>10$ \AA) and higher [\ion{N}{ii}]/H$\alpha$ ratios (log([\ion{N}{ii}]/H$\alpha$) $\gtrsim-0.4$), consistently with the AGN classification by \citet{2011MNRAS.413.1687C}. The few spaxels with slightly lower [\ion{N}{ii}]/H$\alpha$ ratios have however too large \ion{He}{ii}/H$\beta$ ratios to be due to pure star formation.

The right panel (i.e. \ion{He}{ii}/H$\beta$ against EW([\ion{O}{iii}])) highlights the existence of a separate population of spaxels at lower values of EW([\ion{O}{iii}]). Such a spaxel population might either identify cases of AGN nuclear continuum detected to some extent, or central spaxels showing some scattered nuclear emission. We will further investigate this aspect in a future study.

\section{Properties of the AGN host galaxies}
\label{sec:host_properties}
\begin{figure*}
	\includegraphics[width=1.9\columnwidth]{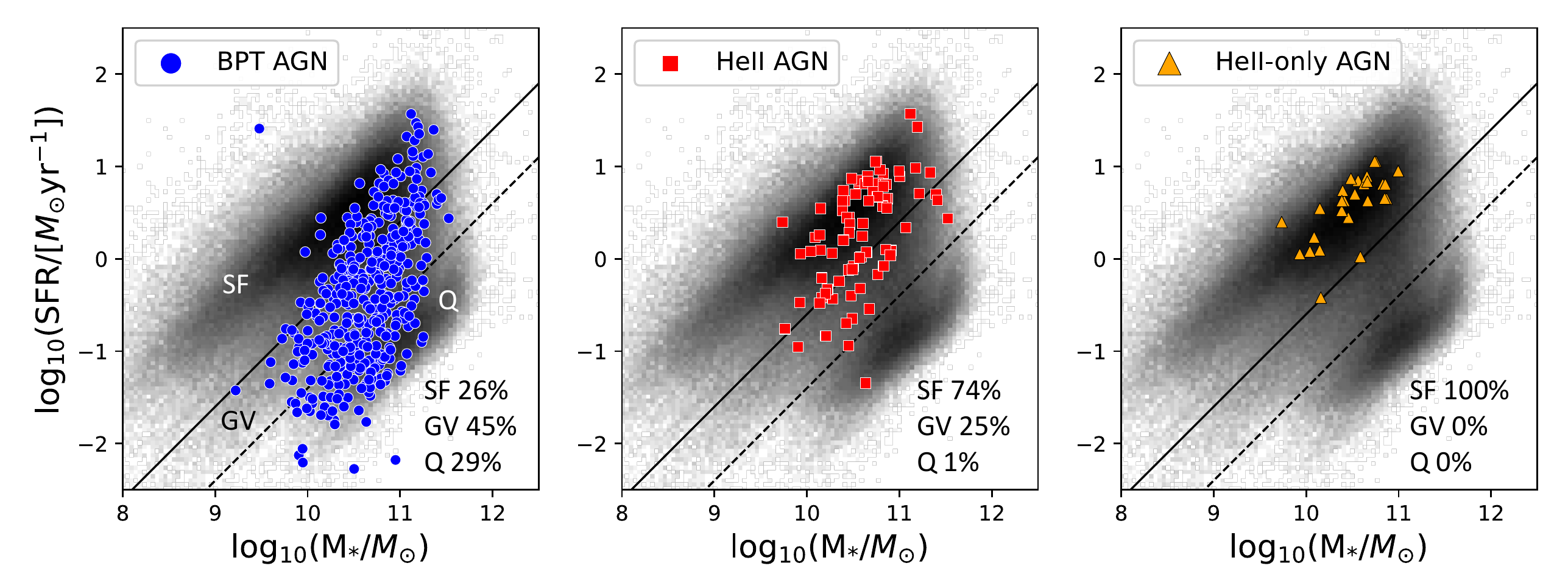}
    \caption{SFR -- M$_*$ diagram separately for the different AGN populations: the BPT-selected AGN (left, blue points), the \ion{He}{ii}-selected AGN (middle, red squares) and the \ion{He}{ii}-only AGN (right, orange triangles). The gray shading shows the distribution of SDSS galaxies and the demarcation lines separating SF, GV and Q galaxies are taken from \citet{2017ApJ...851L..24H}. Each panel reports the AGN percentage lying in the three distinct regions. While BPT-selected AGN mostly fall in the GV as found by previous studies, the \ion{He}{ii}-selected AGN are on the Star Forming Main Sequence, with the \ion{He}{ii}-only AGN exclusively residing on the Star Forming Main Sequence.}
    \label{fig6}
\end{figure*}

\begin{figure}
	\includegraphics[width=0.95\columnwidth]{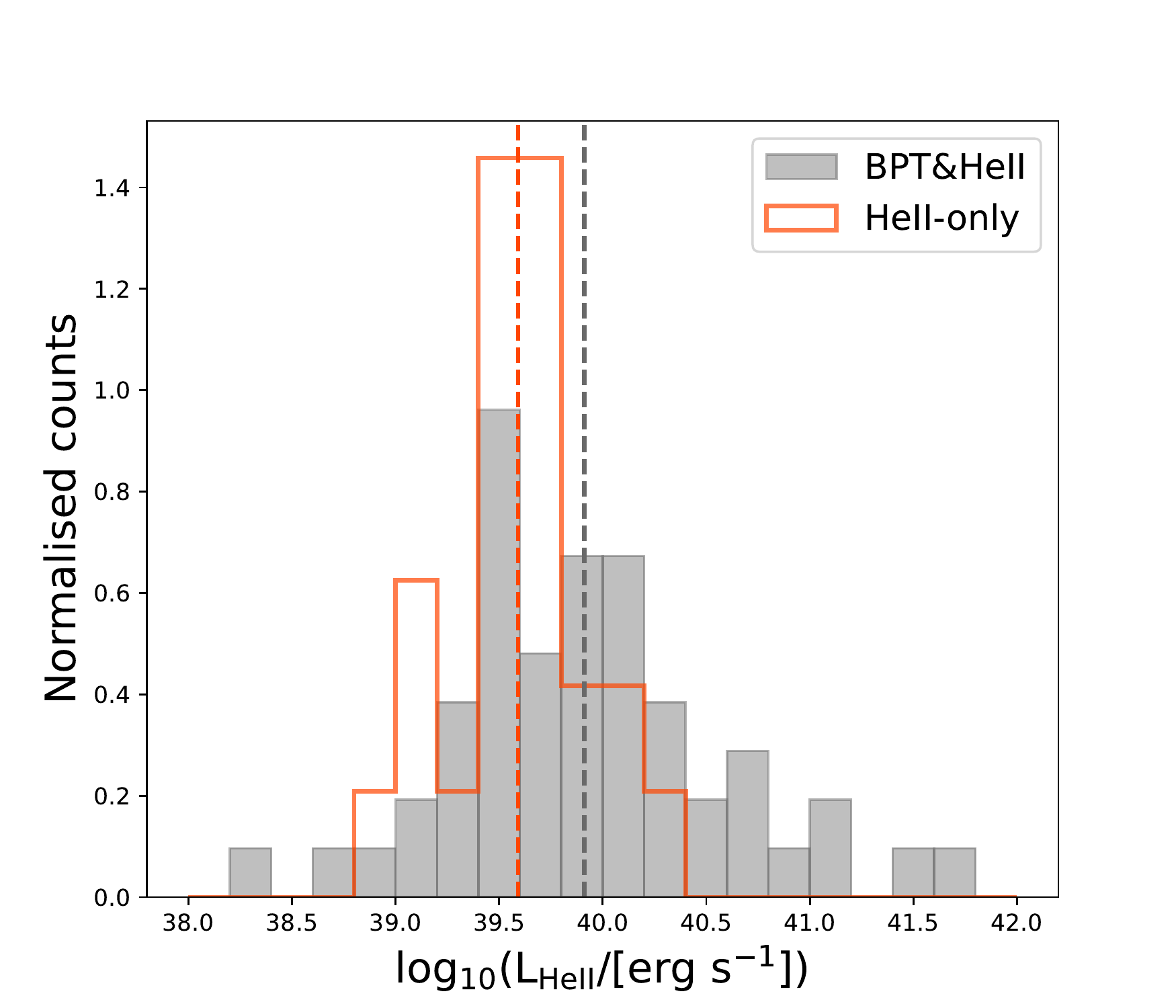}
    \caption{Normalised histogram of the \ion{He}{ii} line luminosity in the BPT\&\ion{He}{ii} (grey-shaded) and \ion{He}{ii}-only AGN (orange line). The resulting mean \ion{He}{ii} luminosity of the two AGN populations are respectively of 39.9 (dashed grey line) and 39.6 (dashed orange) in logarithmic scale.}
    \label{fig7}
\end{figure}

In Fig. \ref{fig6} we plot the star formation rate -- stellar mass diagram (SFR -- M$_*$; \citealt{2007ApJ...660L..43N, 2011A&A...533A.119E, 2015ApJ...801...80L}), separately for the distinct AGN sub-populations, namely the BPT AGN (left panel, blue points), the \ion{He}{ii} AGN (middle, red squares) and the \ion{He}{ii}-only AGN (right, orange triangles) samples, with the grey shading in the background showing the distribution of SDSS galaxies, whose SFR and M$_*$ measurements are taken from the MPA/JHU
catalogue\footnote{Available at http://www.mpa-garching.mpg.de/SDSS/DR7/.}. To separate the different regions in the plane of SF, green valley (GV) and quiescent (Q) galaxies, we use the demarcation lines from \citet{2017ApJ...851L..24H}. Measurements of SFR and M$_*$ have been taken from the MaNGA Pipe3D catalogue. In particular, we adopt the integrated SFR estimates derived from single stellar population (SSP) models provided by Pipe3D, which are computed from the amount of stellar mass formed in the last 32 Myr. Since we are dealing with AGN galaxies, SSP-based SFRs are indeed expected to be more accurate than H$\alpha$-based SFRs (SSP- and H$\alpha$-based SFR values are consistent within 50\% for SF galaxies), given the likely significant AGN contamination to the H$\alpha$ emission.

Each panel of Fig. \ref{fig6} reports the percentage of AGN galaxies of the respective sub-sample lying in the SF, GV and Q regions. The BPT AGN mainly fall in the GV (45\%), with a considerable fraction (29\%) of AGN residing in quiescent galaxies. The \ion{He}{ii} AGN instead tend to lie on the Main Sequence (MS, 74\%), although they are still scattered across the GV (25\%). Finally, the \ion{He}{ii}-only AGN exhibit the interesting property of being all in MS galaxies. All these findings are consistent with those obtained by \citet{2017MNRAS.466.2879B} using single-fiber SDSS measurements. However, it is important to note that our selected \ion{He}{ii}-only AGN reside in massive galaxies (M$_*\gtrsim10^{10}$ \(M_\odot\)). This gives us confidence that we are not including star forming dwarf galaxies, where the AGN presence may be more questionable and the \ion{He}{ii} emission may originate from different sources, such as hot massive stars in metal-poor systems \citep[e.g.][]{2022ApJ...930...37U}.

The discovery through the \ion{He}{ii} diagnostics of an AGN sub-population residing in massive MS galaxies and missed by the BPT classification points to the unique ability of the \ion{He}{ii} emission line diagnostic to find elusive AGN hosted by actively SF galaxies. This hence highlights the possible relevant role of the \ion{He}{ii} line emission in the systematic rest-frame optical search for AGN especially at very high redshift, where the star formation in MS galaxies is expected to be higher.

In Fig. \ref{fig7} we compare the \ion{He}{ii} line luminosity ($L_{\text{\ion{He}{ii}}}$) in \ion{He}{ii}-only AGN to that of BPT\&\ion{He}{ii} AGN selected by both diagnostics. The \ion{He}{ii} line luminosity (as well as other commonly used lines, such as [\ion{O}{iii}]) is indeed known to correlate with the AGN intrinsic luminosity as traced, for instance, by their hard X-ray emission \citep{2015MNRAS.454.3622B}. Therefore, it can be used as a proxy for the AGN luminosity. For this purpose, we compute the total $L_{\text{\ion{He}{ii}}}$ in a given galaxy from the total \ion{He}{ii} line flux obtained by summing the \ion{He}{ii} flux contained in S/N(\ion{He}{ii})$>$5 spaxels. Similarly to what obtained in \citet{2017MNRAS.466.2879B}, we find that the \ion{He}{ii}-only AGN are less luminous than the BPT\&\ion{He}{ii} ones, with a mean \ion{He}{ii} luminosity (in units of erg s$^{-1}$) of 39.6 (dashed orange line) and 39.9 (dashed black line) in logarithmic scale, respectively. The mean values inferred in \citet{2017MNRAS.466.2879B} for SDSS galaxies are larger (40.7 and 41.1, respectively) but also separated by a comparable difference of a few dex. This difference in $L_{\text{\ion{He}{ii}}}$ is likely due to the fact that the SDSS survey spans a higher redshift range and therefore includes, on average, more luminous AGN than MaNGA. 

\section{Discussion}
\label{sec:discussion}

\subsection{Estimating the number of undetected \ion{He}{ii}-only AGN}
\label{sec:missing}

\begin{figure}
	\includegraphics[width=0.95\columnwidth]{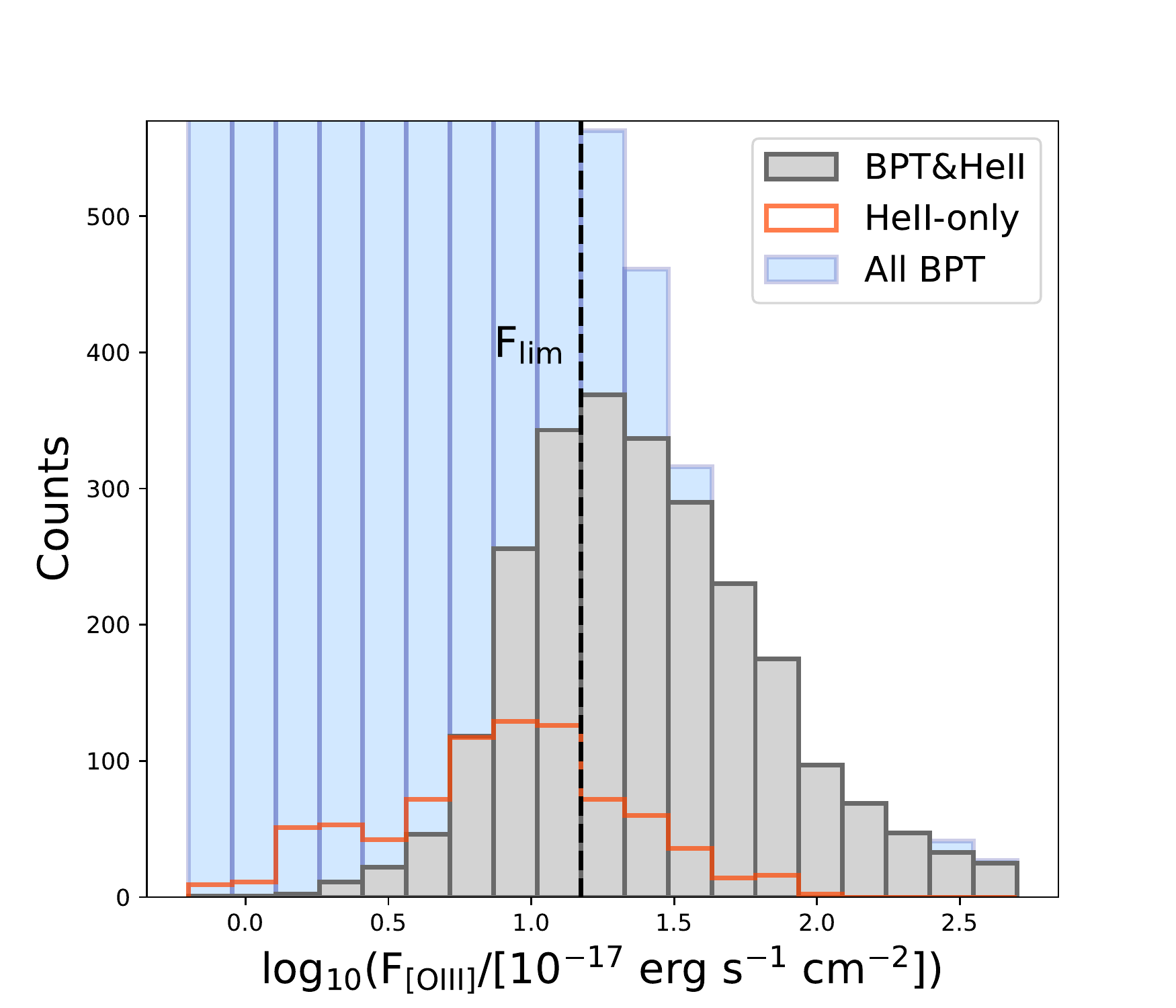}
    \caption{Distribution of the [\ion{O}{iii}] flux (estimated from the total flux of spaxels at radii $R<0.6$ R$_\mathrm{e}$) for the different AGN subsamples: both BPT\&\ion{He}{ii} classified AGN (filled grey
    ),
    \ion{He}{ii}-only classified AGN (orange contours;
    ), and all BPT-classified AGN  (filled lightblue
    ).
    The vertical, black dashed line represents, for the BPT-classified AGN, the [\ion{O}{iii}] flux adopted as detection limit ($F^{\rm lim}_{\rm [\ion{O}{iii}]}\sim1.5\times10^{-16}$ erg s$^{-1}$ cm$^{-2}$) for them to also have an \ion{He}{ii} detection and AGN classification.
    }
    \label{fig8}
\end{figure}

Previously, we showed that there are 27 galaxies (i.e. 6\% of the total selected AGN sample) classified as AGN  only on the basis of the \ion{He}{ii} diagram. Such \ion{He}{ii}-only AGN represent about 33\% of the overall \ion{He}{ii}-selected AGN population and, interestingly, reside primarily in massive (M$_*\gtrsim10^{10}$ \(M_\odot\)) MS galaxies, where SF activity plays a primary role as an ionising source. The intense star formation in the galaxy, combined with the low-luminosity nature of these AGN, can therefore lead to the missed BPT-AGN selection. Being more sensitive to AGN emission, the \ion{He}{ii} diagnostics allows us to unveil this low-luminosity AGN sub-population and to perform a more complete census of the total AGN population.

However, as already mentioned, the main issue related to the use of the \ion{He}{ii} diagnostics is the faintness of such an emission line. Among our total selected AGN population (459), we indeed detect the \ion{He}{ii} line in 81 objects only. It is therefore important to estimate how many \ion{He}{ii} AGN we are missing in the MaNGA sample because of detection limits, so to obtain a less biased estimate of the AGN population. A fraction of this overall undetected \ion{He}{ii} AGN population is expected to be among the selected BPT AGN which show no \ion{He}{ii} line emission (cerulean region in Fig. \ref{fig3}, 378 objects). Therefore, these are already counted in the total AGN census, whereas we want to estimate the number of missing \ion{He}{ii}-only AGN in MaNGA, not yet included in our total selected AGN sample.

With this aim, we first estimate the MaNGA detection limit for the \ion{He}{ii} line emission by using the brighter [\ion{O}{iii}] line emission as a proxy.
Figure \ref{fig8} shows the distribution of the [\ion{O}{iii}] emission line flux contained in single spaxels within a central aperture of radius of 0.6 R$_\mathrm{e}$, separately, for the AGN subsample selected by both diagrams (filled grey histogram; S/N([\ion{O}{iii}])$>$3 and S/N(\ion{He}{ii})$>$5), the \ion{He}{ii}-only AGN population (orange contours; S/N([\ion{O}{iii}])$>$3 and S/N(\ion{He}{ii})$>$5), and the overall BPT-selected AGN (filled lightblue; S/N([\ion{O}{iii}])$>$3). By comparing the \ion{He}{ii}-only and BPT\&\ion{He}{ii} distributions, the former is shifted towards smaller values of [\ion{O}{iii}] flux compared to the latter. This behaviour is expected since in the \ion{He}{ii}-only AGN the star formation dominates over the AGN, thus producing smaller [\ion{O}{iii}]/H$\beta$ ratios consistent with ionisation due to star formation. The BPT\&\ion{He}{ii} is coincident with the overall BPT AGN selected population above
$\rm log(F_{\rm [\ion{O}{iii}]}/10^{-17}erg~s^{-1})>1.5$. This is interesting, as it indicates that at high [\ion{O}{iii}] fluxes all BPT-selected AGN are also \ion{He}{ii}-selected. However, at lower fluxes the two populations start to depart and the BPT\&\ion{He}{ii} distribution shows a clear cutoff at a flux of $F^{\rm lim}_{\rm [\ion{O}{iii}]}<1.5\times10^{-16}$ erg s$^{-1}$ cm$^{-2}$ (vertical black dashed line), while the distribution of all standard BPT AGN keeps increasing to lower [\ion{O}{iii}] fluxes.
The cutoff in the BPT\&\ion{He}{ii} population is clearly due to the requirement of having \ion{He}{ii} detected. Therefore, we take the cutoff value ($1.5\times10^{-16}$ erg s$^{-1}$ cm$^{-2}$) as the limiting [\ion{O}{iii}] flux ($F_{\rm lim}$) beneath which the \ion{He}{ii} emission is likely missed because of the sensitivity of MaNGA observations.

\begin{table}
\centering
\setlength{\tabcolsep}{5pt} 
\renewcommand{\arraystretch}{1} 
\caption{Demography of AGN galaxies included in the BPT-selected (first row) and \ion{He}{ii}-only-selected (second row) AGN subsamples. For each subsample, we report the total number of detected objects, and among these the number below ($N_{\mathrm{below-F_{lim}}}$) and above ($N_{\mathrm{above-F_{lim}}}$) our adopted detection limit ($F_{\rm lim}= 1.5\times10^{-16}$ erg s$^{-1}$ cm$^{-2}$. We show the total BPT-selected AGN population as sum of two distinct contributions, namely, the BPT\&\ion{He}{ii} AGN sample and the BPT AGN with no \ion{He}{ii} detection (no\ion{He}{ii}).}
\begin{tabular}{ccccc}
\hline
    & Total &  \multirow{2}{*}{-} &  \multirow{2}{*}{$N_{\mathrm{below-F_{lim}}}$} &  \multirow{2}{*}{$N_{\mathrm{above-F_{lim}}}$}\\
    & detected & & & \\
 \hline
 \hline
 \multirow{2}{*}{\textbf{BPT AGN}} & \multirow{2}{*}{432} & BPT\&\ion{He}{ii} & 29 & 25\\ 
  &  & no\ion{He}{ii} & 376 & 2\\ 
 \hline
 \textbf{\ion{He}{ii}-only AGN} & 27 & - & 22 & 5\\ 
   \hline
\end{tabular}
\label{tab2}
\end{table}

To count how many \ion{He}{ii}-only AGN are missed below this limit, we assume that the ratio of \ion{He}{ii}-only to all BPT AGN below the [\ion{O}{iii}] flux limit is the same as above the limit, namely:



\begin{equation}
\centering
   N^{\mathrm{\ion{He}{ii}-only}}_{\mathrm{below-F_{lim}}} = \frac{N^{\mathrm{\ion{He}{ii}-only}}_{\mathrm{above-F_{lim}}}}{N^{\mathrm{BPT}}_{\mathrm{above-F_{lim}}}}\times N^{\mathrm{BPT}}_{\mathrm{below-F_{lim}}}=75,
\end{equation}
where $N^{\mathrm{\ion{He}{ii}-only}}_{\mathrm{above-F_{lim}}}=5$, $N^{\mathrm{BPT}}_{\mathrm{above-F_{lim}}}=27$, and $N^{\mathrm{BPT}}_{\mathrm{below-F_{lim}}}=405$, as inferred by comparing for each object the average [\ion{O}{iii}] flux in spaxels within 0.6 R$_{\rm e}$ with $F_{\rm lim}$. The resulting demography is summarised in Table~\ref{tab2}. Among the 75 \ion{He}{ii}-only AGN below our adopted limit, 22 are however detected. Therefore, the undetected \ion{He}{ii}-only AGN below $F_{\rm lim}$ are expected to be 53 in total, thus leading to a final overall AGN sample consisting of 512 galaxies. However, we argue that the estimate of 53 missing \ion{He}{ii}-only AGN must be considered as a lower limit to the real number, since such value has been inferred under the assumption that the ratio of \ion{He}{ii}-only AGN to BPT AGN is the same at all stellar masses. Instead, we know that BPT AGN mainly reside in GV and Q galaxies, which are typically more massive than SF Main Sequence galaxies, the primary hosts of \ion{He}{ii}-only AGN.

\subsection{Implications for the AGN census and quenching scenarios}

\begin{figure}
	\includegraphics[width=0.95\columnwidth]{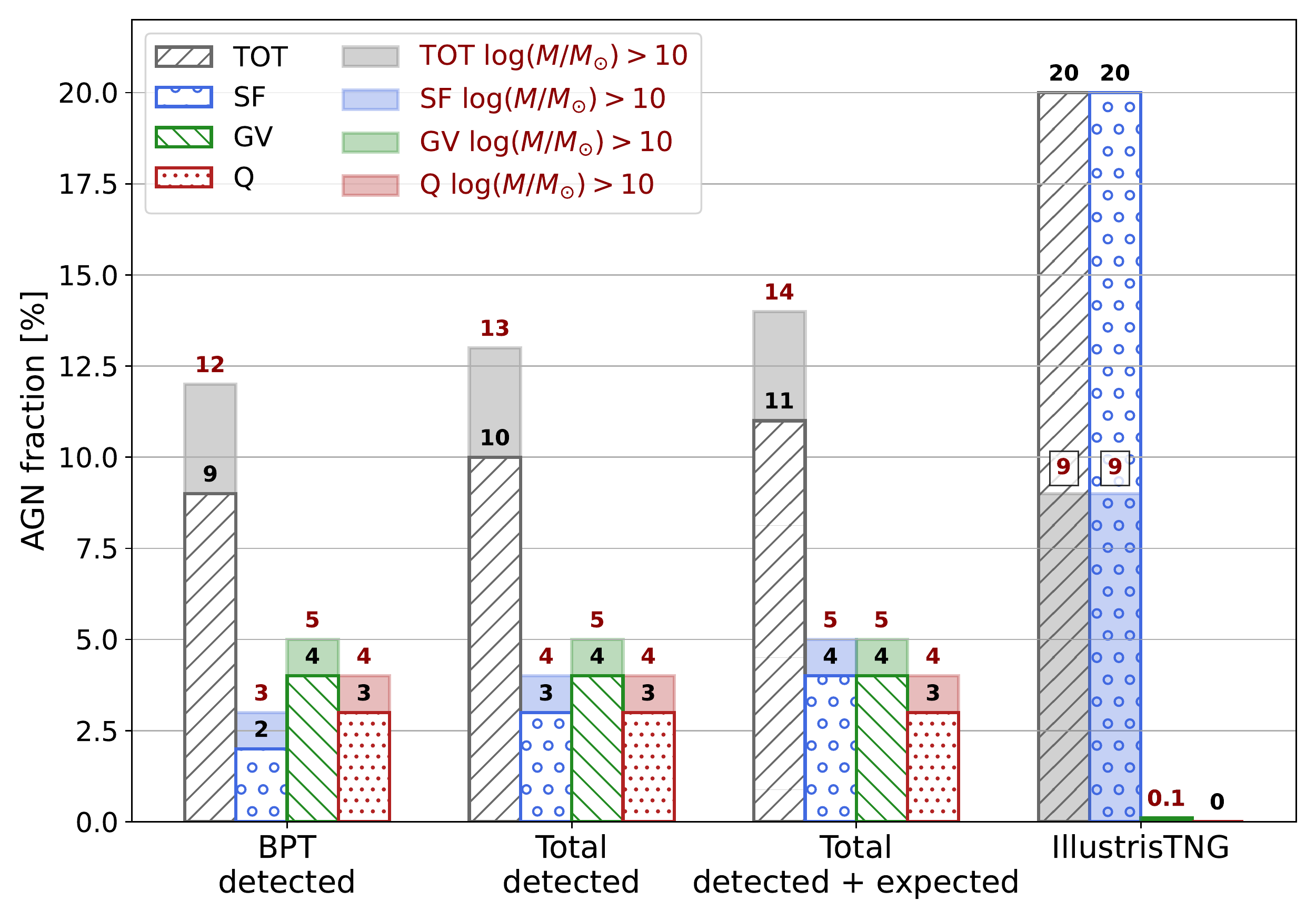}
    \caption{Bar charts summarising the demography in the SFR -- M$_*$ plane of the AGN samples selected by just the BPT (first bar chart), by combining the BPT and \ion{He}{ii} diagnostics (second), and after including the AGN population expected to be missing (53) in the \ion{He}{ii}-only AGN sample because of sensitivity limits (third); finally, the predictions from the IllustrisTNG cosmological simulation (fourth). Each grouped bar chart shows percent fractions, computed with respect to the overall MaNGA sample, of the total AGN (grey) population and, separately, of the AGN residing in SF (blue), GV (green) and Q (red) galaxies. The filled bars (with red labels) represent the same categories but only for the galaxies more massive than M$_*>10^{10}$ \(M_\odot\).}
    \label{fig9}
\end{figure}

As final step of this work, in Fig. \ref{fig9} we study how the AGN demography in the SFR -- M$_*$ plane changes with respect to the initial BPT-based census (first bar chart), after the inclusion of the \ion{He}{ii}-selected AGN (second), and further after the inclusion of the 53 estimated missing \ion{He}{ii}-only AGN (third). Each bar chart shows percent fraction of the total AGN population (grey) and, separately, of the SF Main Sequence (blue), GV (green) and Q (red) AGN sub-samples, with respect to the total MaNGA DR15 galaxies (4656). The filled bars represent the same categories but considering only galaxies more massive than M$_*\gtrsim10^{10}$ \(M_\odot\) (3297 massive galaxies in MaNGA DR15), since almost all (93\%) \ion{He}{ii}-only AGN are found in this stellar mass range. With respect to our initial BPT AGN census, the \ion{He}{ii} diagnostics leads to an increase by 6\% and 24\% in our total and SF Main Sequence detected AGN samples, respectively, further highlighting the poor BPT efficiency at unveiling AGN in SF galaxies. On the contrary, the GV (4\%) and Q (3\%) AGN fractions in MaNGA are not affected at all, being all the selected \ion{He}{ii}-only AGN located on the MS. We assumed the same to be valid for all the 53 \ion{He}{ii}-only AGN expected to lie below the MaNGA detection limit. The inclusion of these 53 objects entails an increase by about 12\% in our total (BPT and/or \ion{He}{ii}) AGN sample, and by about 38\% in our SF (Main Sequence) AGN population.
These expected new figures represent 11\% and 4\% in the overall MaNGA DR15 sample, respectively, corresponding to 14\% and 5\% for the more massive case (M$_*\gtrsim10^{10}$ \(M_\odot\)).

Overall our new census implies a modest increase in the overall census of AGN with respect to the classical BPT classification (from 9\% to 11\%), but the number of AGN on the SF Main Sequence is {\it doubled}.

A substantial number of AGN in SF galaxies compared to GV and Q galaxies points to a picture where significant star formation and black hole accretion are coeval, with a large amount of gas simultaneously available for both star formation and black hole feeding. Such a correlation between star formation and black hole accretion rates is also predicted by cosmological hydrodynamical simulations \citep{2022MNRAS.512.1052P, 2022MNRAS.514.2936W}, such as IllustrisTNG \citep{2018MNRAS.480.5113M, 2018MNRAS.477.1206N, 2018MNRAS.475..624N, 2018MNRAS.475..648P, 2018MNRAS.475..676S}, SIMBA \citep{2019MNRAS.486.2827D} and, to a lesser extent, by EAGLE \citep{2015MNRAS.446..521S}.

To illustrate this in our context, in Fig. \ref{fig9} we also show the predictions of IllustrisTNG for the redshift z$=$0 AGN population (fourth bar chart) for comparison with our observational results. In the simulation, we select as AGN hosts those galaxies with an Eddington ratio $\lambda_{\rm Edd}$ in the top 20\% of all Eddington ratios ($\lambda_{\rm Edd}>0.016$), and classify them in SF, GV and Q galaxies according to the same prescription used with MaNGA data (i.e. \citealt{2017ApJ...851L..24H}). When comparing our observational results (first to third bar charts) with the simulation predictions, we can clearly see that the discrepancy in the fraction of total and SF (MS) AGN starts is mitigated only when adding the  \ion{He}{ii}-only AGN to the BPT-selected sample.
At the same time, we note that the AGN feedback prescription implemented in IllustrisTNG necessarily dictates an association between high AGN accretion rates and their location on the star forming Main Sequence. In this model, AGN feedback is only successful at suppressing global SFR in galaxies when it switches into the `kinetic' mode at low accretion rates. In consequence, simulated black holes with high $\lambda_{\rm Edd}$ are expected to almost exclusively reside in galaxies with active star formation.



Finally, we highlight that the existence of a considerable AGN population in SF galaxies compared to GV and Q galaxies can have important implications for AGN quenching scenarios \citep{2009ApJ...692L..19S}.
In fact, the remarkably larger fraction of AGN found in GV and Q galaxies in past studies has always favoured either the scenario in which the AGN phase is long-lived enough ($\sim$1 Gyr) to suppress the SF, and then evolve towards the green valley and the red sequence, or the scenario in which AGN are able to immediately quench the SF but becoming detectable only with a delay of about 100 Myr. Our discovery of a substantial AGN population in SF galaxies, instead might suggest the scenario in which the AGN does not really quench star formation directly nor instantaneously, but suppress start formation indirectly, on much longer timescales ($\sim$1~Gyr) via `preventative' feedback \citep{2012ARA&A..50..455F, 2020MNRAS.499..230B, 2020MNRAS.491.5406T, 2022MNRAS.512.1052P}, namely, by injecting energy into the halo, preventing accretion of fresh gas, and hampering star formation as a consequence of starvation.





\section{Conclusions}
\label{sec:conclusion}
We have used the \ion{He}{ii} diagnostics \citep{2012MNRAS.421.1043S} to detect AGN activity in galaxies observed in the MaNGA survey, hence with spatially-resolved spectroscopy, and compare the resulting population with the AGN detected via the standard BPT diagrams. The main conclusions from our study are summarised below.
\begin{enumerate}
    \item By comparing the spatially-resolved BPT and \ion{He}{ii} diagrams of MaNGA spaxels (Fig. \ref{fig1}), we find that whereas the \ion{He}{ii} diagnostics globally recover the BPT AGN classification, only 29\% of the \ion{He}{ii}-selected AGN spaxels are classified as AGN by the BPT diagram.
    \item Based on the spaxel classification according to the two diagnostics, we obtain a total sample of 459 AGN host galaxies in MaNGA, out of which 432 (94\%) are BPT-selected AGN, whereas 81 (18\%) are \ion{He}{ii}-selected AGN (Fig. \ref{fig3}). In the \ion{He}{ii} AGN sample, 27 objects (6\%) are classified as AGN by the \ion{He}{ii} diagnostics only (\ion{He}{ii}-only AGN), out of which 11 (41\%) are identified as AGN based on radio observations (10) and/or mid-IR colours (2) \citep{2020ApJ...901..159C}. This further supports the \ion{He}{ii} diagnostic as useful tracer of AGN missed in usual BPT diagnostics.
    \item The \ion{He}{ii} AGN-like emission is overall centrally located (within 0.6 R$_\mathrm{e}$) in all \ion{He}{ii}-selected AGN galaxies (Fig. \ref{fig4}), thus supporting the AGN-like ionisation of such \ion{He}{ii} emitting regions. Moreover, the high \ion{He}{ii}/H$\beta$ and [\ion{N}{ii}]/H$\alpha$ ratios combined with large EWs (EW(H$\alpha$) $>10$ \AA) (Fig. \ref{fig5}), compared to pure SF galaxies \citep{2011MNRAS.413.1687C}, further supports our correct identification of the 27 \ion{He}{ii}-only AGN as real active galaxies.
    \item All \ion{He}{ii}-only AGN reside in Star Forming Main Sequence galaxies (Fig. \ref{fig6}) and are less luminous than the BPT-selected AGN, on average (Fig. \ref{fig7}). These lower-luminosity AGN are overwhelmed by the intense star formation radiation and, therefore, miss the BPT AGN classification. Furthermore, the \ion{He}{ii}-only AGN are mainly found in massive galaxies (M$_*\gtrsim10^{10}$ \(M_\odot\)). This rules out the possibility that we are including SF dwarf galaxies, whose \ion{He}{ii} emission might be due to hot massive stars and X-ray binaries, especially in case of metal-poor environments \citep{2022ApJ...930...37U}.
    \item Being the \ion{He}{ii} an extremely faint emission line, we expect several more AGN to be below the \ion{He}{ii} detection limit in MaNGA (Fig. \ref{fig8}). Under simplifying assumptions, we estimated a lower limit to the missing population of undetected \ion{He}{ii}-only AGN (53 objects), all assumed to reside in SF galaxies like the objects we detect.
    \item Our inferred, revised AGN census indicates that 11\% of galaxies (in MaNGA DR15) host an AGN, of which 4\% on the SF Main Sequence, 4\% in the Green Valley and 3\% in Quiescent galaxies (Fig. \ref{fig9}). We note that on the SF Main Sequence the number of AGN is doubled with respect to the simple BPT classification. If we restrict the census to galaxies more massive than M$_*\gtrsim10^{10}$ \(M_\odot\), then we infer that the AGN population rises to 14\%, of which 5\% on the Star Forming Main Sequence, with the the Green Valley (5\%) and Quiescent (4\%) AGN populations unaffected.
    \item The presence of a substantial number of AGN in SF galaxies compared to GV and Q galaxies is consistent with expectations of cosmological simulations \citep{2022MNRAS.512.1052P, 2022MNRAS.514.2936W}. It may also have important implications on quenching scenarios, possibly supporting the picture where AGN feedback on SF is not instantaneous (i.e. inefficiency of the ejective mode) but, if anything, has a delayed, preventive quenching effect, likely via halo heating.
\end{enumerate}

With a view to exploring high redshifts at unprecedented high resolution and sensitivity with JWST and upcoming VLT facilities (ERIS and MOONS), novel emission-line diagnostics will be fundamental to investigate the ionisation conditions of the high-redshift Universe, featured by highly SF and/or metal-poor environments. As photoionisation models predict BPT to fail in correctly identifying the dominant ionisation process in case of metal-poor conditions \citep{2022MNRAS.513.5134N}, our study has demonstrated the crucial role the \ion{He}{ii} diagnostic might play in the search for hidden AGN in actively SF galaxies. We finally point out the greater feasibility of the \ion{He}{ii} diagnostics compared to other BPT-alternative AGN tracers such as the rarer and fainter optical coronal lines (e.g. [\ion{Ar}{X}]$\lambda$5533, [\ion{S}{XII}]$\lambda$7609, [\ion{Fe}{XI}]$\lambda$7982, [\ion{Fe}{XIV}]$\lambda$5303; \citealt{2009MNRAS.397..172G,2021ApJ...922..155M,2021ApJ...920...62N,2023arXiv230113322N,2022ApJ...936..140R}).


\section*{Acknowledgements}
G.T., R.M., J.M.P. and M.C. acknowledge funding from ERC Advanced Grant 695671 ``QUENCH’’, as well as support by the Science and Technology Facilities Council (STFC). R.M. additionally acknowledges the support from a Royal Society Research Professorship. Finally, we are grateful to J. Brinchmann for providing us the equation of the 50\% AGN demarcation line to show in the \ion{He}{ii} diagram.

\section*{Data Availability}
The MaNGA data used in this work are publicly available at https://www.sdss.org/dr15/manga/manga-data/.




\bibliographystyle{mnras}
\bibliography{mybiblio} 

\begin{thebibliography}{}
\makeatletter
\relax
\def\mn@urlcharsother{\let\do\@makeother \do\$\do\&\do\#\do\^\do\_\do\%\do\~}
\def\mn@doi{\begingroup\mn@urlcharsother \@ifnextchar [ {\mn@doi@}
  {\mn@doi@[]}}
\def\mn@doi@[#1]#2{\def\@tempa{#1}\ifx\@tempa\@empty \href
  {http://dx.doi.org/#2} {doi:#2}\else \href {http://dx.doi.org/#2} {#1}\fi
  \endgroup}
\def\mn@eprint#1#2{\mn@eprint@#1:#2::\@nil}
\def\mn@eprint@arXiv#1{\href {http://arxiv.org/abs/#1} {{\tt arXiv:#1}}}
\def\mn@eprint@dblp#1{\href {http://dblp.uni-trier.de/rec/bibtex/#1.xml}
  {dblp:#1}}
\def\mn@eprint@#1:#2:#3:#4\@nil{\def\@tempa {#1}\def\@tempb {#2}\def\@tempc
  {#3}\ifx \@tempc \@empty \let \@tempc \@tempb \let \@tempb \@tempa \fi \ifx
  \@tempb \@empty \def\@tempb {arXiv}\fi \@ifundefined
  {mn@eprint@\@tempb}{\@tempb:\@tempc}{\expandafter \expandafter \csname
  mn@eprint@\@tempb\endcsname \expandafter{\@tempc}}}

\bibitem[\protect\citeauthoryear{{Abazajian} et~al.,}{{Abazajian}
  et~al.}{2009}]{2009ApJS..182..543A}
{Abazajian} K.~N.,  et~al., 2009, \mn@doi [\apjs]
  {10.1088/0067-0049/182/2/543}, \href
  {https://ui.adsabs.harvard.edu/abs/2009ApJS..182..543A} {182, 543}

\bibitem[\protect\citeauthoryear{{Ajello}, {Alexander}, {Greiner}, {Madejski},
  {Gehrels}  \& {Burlon}}{{Ajello} et~al.}{2012}]{2012ApJ...749...21A}
{Ajello} M.,  {Alexander} D.~M.,  {Greiner} J.,  {Madejski} G.~M.,  {Gehrels}
  N.,   {Burlon} D.,  2012, \mn@doi [\apj] {10.1088/0004-637X/749/1/21}, \href
  {https://ui.adsabs.harvard.edu/abs/2012ApJ...749...21A} {749, 21}

\bibitem[\protect\citeauthoryear{{Baldwin}, {Phillips}  \&
  {Terlevich}}{{Baldwin} et~al.}{1981}]{1981PASP...93....5B}
{Baldwin} J.~A.,  {Phillips} M.~M.,   {Terlevich} R.,  1981, \mn@doi [\pasp]
  {10.1086/130766}, \href
  {https://ui.adsabs.harvard.edu/abs/1981PASP...93....5B} {93, 5}

\bibitem[\protect\citeauthoryear{{B{\"a}r}, {Weigel}, {Sartori}, {Oh}, {Koss}
  \& {Schawinski}}{{B{\"a}r} et~al.}{2017}]{2017MNRAS.466.2879B}
{B{\"a}r} R.~E.,  {Weigel} A.~K.,  {Sartori} L.~F.,  {Oh} K.,  {Koss} M.,
  {Schawinski} K.,  2017, \mn@doi [\mnras] {10.1093/mnras/stw3283}, \href
  {https://ui.adsabs.harvard.edu/abs/2017MNRAS.466.2879B} {466, 2879}

\bibitem[\protect\citeauthoryear{{Belfiore} et~al.,}{{Belfiore}
  et~al.}{2019}]{2019AJ....158..160B}
{Belfiore} F.,  et~al., 2019, \mn@doi [\aj] {10.3847/1538-3881/ab3e4e}, \href
  {https://ui.adsabs.harvard.edu/abs/2019AJ....158..160B} {158, 160}

\bibitem[\protect\citeauthoryear{{Benson}, {Bower}, {Frenk}, {Lacey}, {Baugh}
  \& {Cole}}{{Benson} et~al.}{2003}]{2003ApJ...599...38B}
{Benson} A.~J.,  {Bower} R.~G.,  {Frenk} C.~S.,  {Lacey} C.~G.,  {Baugh} C.~M.,
    {Cole} S.,  2003, \mn@doi [\apj] {10.1086/379160}, \href
  {https://ui.adsabs.harvard.edu/abs/2003ApJ...599...38B} {599, 38}

\bibitem[\protect\citeauthoryear{{Berney} et~al.,}{{Berney}
  et~al.}{2015}]{2015MNRAS.454.3622B}
{Berney} S.,  et~al., 2015, \mn@doi [\mnras] {10.1093/mnras/stv2181}, \href
  {https://ui.adsabs.harvard.edu/abs/2015MNRAS.454.3622B} {454, 3622}

\bibitem[\protect\citeauthoryear{{Blanton} et~al.,}{{Blanton}
  et~al.}{2017}]{2017AJ....154...28B}
{Blanton} M.~R.,  et~al., 2017, \mn@doi [\aj] {10.3847/1538-3881/aa7567}, \href
  {https://ui.adsabs.harvard.edu/abs/2017AJ....154...28B} {154, 28}

\bibitem[\protect\citeauthoryear{{Bluck} et~al.,}{{Bluck}
  et~al.}{2020}]{2020MNRAS.499..230B}
{Bluck} A. F.~L.,  et~al., 2020, \mn@doi [\mnras] {10.1093/mnras/staa2806},
  \href {https://ui.adsabs.harvard.edu/abs/2020MNRAS.499..230B} {499, 230}

\bibitem[\protect\citeauthoryear{{Brinchmann}, {Kunth}  \&
  {Durret}}{{Brinchmann} et~al.}{2008}]{2008A&A...485..657B}
{Brinchmann} J.,  {Kunth} D.,   {Durret} F.,  2008, \mn@doi [\aap]
  {10.1051/0004-6361:200809783}, \href
  {https://ui.adsabs.harvard.edu/abs/2008A&A...485..657B} {485, 657}

\bibitem[\protect\citeauthoryear{{Bundy} et~al.,}{{Bundy}
  et~al.}{2015}]{2015ApJ...798....7B}
{Bundy} K.,  et~al., 2015, \mn@doi [\apj] {10.1088/0004-637X/798/1/7}, \href
  {https://ui.adsabs.harvard.edu/abs/2015ApJ...798....7B} {798, 7}

\bibitem[\protect\citeauthoryear{{Calzetti}, {Armus}, {Bohlin}, {Kinney},
  {Koornneef}  \& {Storchi-Bergmann}}{{Calzetti}
  et~al.}{2000}]{2000ApJ...533..682C}
{Calzetti} D.,  {Armus} L.,  {Bohlin} R.~C.,  {Kinney} A.~L.,  {Koornneef} J.,
   {Storchi-Bergmann} T.,  2000, \mn@doi [\apj] {10.1086/308692}, \href
  {https://ui.adsabs.harvard.edu/abs/2000ApJ...533..682C} {533, 682}

\bibitem[\protect\citeauthoryear{{Cappellari}}{{Cappellari}}{2017}]{2017MNRAS.466..798C}
{Cappellari} M.,  2017, \mn@doi [\mnras] {10.1093/mnras/stw3020}, \href
  {https://ui.adsabs.harvard.edu/abs/2017MNRAS.466..798C} {466, 798}

\bibitem[\protect\citeauthoryear{{Cappellari} \& {Emsellem}}{{Cappellari} \&
  {Emsellem}}{2004}]{2004PASP..116..138C}
{Cappellari} M.,  {Emsellem} E.,  2004, \mn@doi [\pasp] {10.1086/381875}, \href
  {https://ui.adsabs.harvard.edu/abs/2004PASP..116..138C} {116, 138}

\bibitem[\protect\citeauthoryear{{Cherinka} et~al.,}{{Cherinka}
  et~al.}{2019}]{2019AJ....158...74C}
{Cherinka} B.,  et~al., 2019, \mn@doi [\aj] {10.3847/1538-3881/ab2634}, \href
  {https://ui.adsabs.harvard.edu/abs/2019AJ....158...74C} {158, 74}

\bibitem[\protect\citeauthoryear{{Cid Fernandes}, {Stasi{\'n}ska}, {Mateus}  \&
  {Vale Asari}}{{Cid Fernandes} et~al.}{2011}]{2011MNRAS.413.1687C}
{Cid Fernandes} R.,  {Stasi{\'n}ska} G.,  {Mateus} A.,   {Vale Asari} N.,
  2011, \mn@doi [\mnras] {10.1111/j.1365-2966.2011.18244.x}, \href
  {https://ui.adsabs.harvard.edu/abs/2011MNRAS.413.1687C} {413, 1687}

\bibitem[\protect\citeauthoryear{{Comerford} \& {Greene}}{{Comerford} \&
  {Greene}}{2014}]{2014ApJ...789..112C}
{Comerford} J.~M.,  {Greene} J.~E.,  2014, \mn@doi [\apj]
  {10.1088/0004-637X/789/2/112}, \href
  {https://ui.adsabs.harvard.edu/abs/2014ApJ...789..112C} {789, 112}

\bibitem[\protect\citeauthoryear{{Comerford}, {Gerke}, {Stern}, {Cooper},
  {Weiner}, {Newman}, {Madsen}  \& {Barrows}}{{Comerford}
  et~al.}{2012}]{2012ApJ...753...42C}
{Comerford} J.~M.,  {Gerke} B.~F.,  {Stern} D.,  {Cooper} M.~C.,  {Weiner}
  B.~J.,  {Newman} J.~A.,  {Madsen} K.,   {Barrows} R.~S.,  2012, \mn@doi
  [\apj] {10.1088/0004-637X/753/1/42}, \href
  {https://ui.adsabs.harvard.edu/abs/2012ApJ...753...42C} {753, 42}

\bibitem[\protect\citeauthoryear{{Comerford} et~al.,}{{Comerford}
  et~al.}{2020}]{2020ApJ...901..159C}
{Comerford} J.~M.,  et~al., 2020, \mn@doi [\apj] {10.3847/1538-4357/abb2ae},
  \href {https://ui.adsabs.harvard.edu/abs/2020ApJ...901..159C} {901, 159}

\bibitem[\protect\citeauthoryear{{Cresci} et~al.,}{{Cresci}
  et~al.}{2023}]{2023arXiv230111060C}
{Cresci} G.,  et~al., 2023, \mn@doi [arXiv e-prints]
  {10.48550/arXiv.2301.11060}, \href
  {https://ui.adsabs.harvard.edu/abs/2023arXiv230111060C} {p. arXiv:2301.11060}

\bibitem[\protect\citeauthoryear{{Croton} et~al.,}{{Croton}
  et~al.}{2006}]{2006MNRAS.365...11C}
{Croton} D.~J.,  et~al., 2006, \mn@doi [\mnras]
  {10.1111/j.1365-2966.2005.09675.x}, \href
  {https://ui.adsabs.harvard.edu/abs/2006MNRAS.365...11C} {365, 11}

\bibitem[\protect\citeauthoryear{{Dav{\'e}}, {Angl{\'e}s-Alc{\'a}zar},
  {Narayanan}, {Li}, {Rafieferantsoa}  \& {Appleby}}{{Dav{\'e}}
  et~al.}{2019}]{2019MNRAS.486.2827D}
{Dav{\'e}} R.,  {Angl{\'e}s-Alc{\'a}zar} D.,  {Narayanan} D.,  {Li} Q.,
  {Rafieferantsoa} M.~H.,   {Appleby} S.,  2019, \mn@doi [\mnras]
  {10.1093/mnras/stz937}, \href
  {https://ui.adsabs.harvard.edu/abs/2019MNRAS.486.2827D} {486, 2827}

\bibitem[\protect\citeauthoryear{{Drory} et~al.,}{{Drory}
  et~al.}{2015}]{2015AJ....149...77D}
{Drory} N.,  et~al., 2015, \mn@doi [\aj] {10.1088/0004-6256/149/2/77}, \href
  {https://ui.adsabs.harvard.edu/abs/2015AJ....149...77D} {149, 77}

\bibitem[\protect\citeauthoryear{{Elbaz} et~al.,}{{Elbaz}
  et~al.}{2011}]{2011A&A...533A.119E}
{Elbaz} D.,  et~al., 2011, \mn@doi [\aap] {10.1051/0004-6361/201117239}, \href
  {https://ui.adsabs.harvard.edu/abs/2011A&A...533A.119E} {533, A119}

\bibitem[\protect\citeauthoryear{{Fabian}}{{Fabian}}{2012}]{2012ARA&A..50..455F}
{Fabian} A.~C.,  2012, \mn@doi [\araa] {10.1146/annurev-astro-081811-125521},
  \href {https://ui.adsabs.harvard.edu/abs/2012ARA&A..50..455F} {50, 455}

\bibitem[\protect\citeauthoryear{{Garnett}, {Kennicutt}, {Chu}  \&
  {Skillman}}{{Garnett} et~al.}{1991}]{1991ApJ...373..458G}
{Garnett} D.~R.,  {Kennicutt} Robert~C. J.,  {Chu} Y.-H.,   {Skillman} E.~D.,
  1991, \mn@doi [\apj] {10.1086/170065}, \href
  {https://ui.adsabs.harvard.edu/abs/1991ApJ...373..458G} {373, 458}

\bibitem[\protect\citeauthoryear{{Gelbord}, {Mullaney}  \& {Ward}}{{Gelbord}
  et~al.}{2009}]{2009MNRAS.397..172G}
{Gelbord} J.~M.,  {Mullaney} J.~R.,   {Ward} M.~J.,  2009, \mn@doi [\mnras]
  {10.1111/j.1365-2966.2009.14961.x}, \href
  {https://ui.adsabs.harvard.edu/abs/2009MNRAS.397..172G} {397, 172}

\bibitem[\protect\citeauthoryear{{Greene}, {Zakamska}, {Ho}  \&
  {Barth}}{{Greene} et~al.}{2011}]{2011ApJ...732....9G}
{Greene} J.~E.,  {Zakamska} N.~L.,  {Ho} L.~C.,   {Barth} A.~J.,  2011, \mn@doi
  [\apj] {10.1088/0004-637X/732/1/9}, \href
  {https://ui.adsabs.harvard.edu/abs/2011ApJ...732....9G} {732, 9}

\bibitem[\protect\citeauthoryear{{Gunn} et~al.,}{{Gunn}
  et~al.}{2006}]{2006AJ....131.2332G}
{Gunn} J.~E.,  et~al., 2006, \mn@doi [\aj] {10.1086/500975}, \href
  {https://ui.adsabs.harvard.edu/abs/2006AJ....131.2332G} {131, 2332}

\bibitem[\protect\citeauthoryear{{Hainline}, {Hickox}, {Chen}, {Carroll},
  {Jones}, {Zervos}  \& {Goulding}}{{Hainline}
  et~al.}{2016}]{2016ApJ...823...42H}
{Hainline} K.~N.,  {Hickox} R.~C.,  {Chen} C.-T.,  {Carroll} C.~M.,  {Jones}
  M.~L.,  {Zervos} A.~S.,   {Goulding} A.~D.,  2016, \mn@doi [\apj]
  {10.3847/0004-637X/823/1/42}, \href
  {https://ui.adsabs.harvard.edu/abs/2016ApJ...823...42H} {823, 42}

\bibitem[\protect\citeauthoryear{{Hsieh} et~al.,}{{Hsieh}
  et~al.}{2017}]{2017ApJ...851L..24H}
{Hsieh} B.~C.,  et~al., 2017, \mn@doi [\apjl] {10.3847/2041-8213/aa9d80}, \href
  {https://ui.adsabs.harvard.edu/abs/2017ApJ...851L..24H} {851, L24}

\bibitem[\protect\citeauthoryear{{Kauffmann} et~al.,}{{Kauffmann}
  et~al.}{2003}]{2003MNRAS.346.1055K}
{Kauffmann} G.,  et~al., 2003, \mn@doi [\mnras]
  {10.1111/j.1365-2966.2003.07154.x}, \href
  {https://ui.adsabs.harvard.edu/abs/2003MNRAS.346.1055K} {346, 1055}

\bibitem[\protect\citeauthoryear{{Keel} et~al.,}{{Keel}
  et~al.}{2012}]{2012MNRAS.420..878K}
{Keel} W.~C.,  et~al., 2012, \mn@doi [\mnras]
  {10.1111/j.1365-2966.2011.20101.x}, \href
  {https://ui.adsabs.harvard.edu/abs/2012MNRAS.420..878K} {420, 878}

\bibitem[\protect\citeauthoryear{{Keel} et~al.,}{{Keel}
  et~al.}{2015}]{2015AJ....149..155K}
{Keel} W.~C.,  et~al., 2015, \mn@doi [\aj] {10.1088/0004-6256/149/5/155}, \href
  {https://ui.adsabs.harvard.edu/abs/2015AJ....149..155K} {149, 155}

\bibitem[\protect\citeauthoryear{{Kewley}, {Dopita}, {Sutherland}, {Heisler}
  \& {Trevena}}{{Kewley} et~al.}{2001}]{2001ApJ...556..121K}
{Kewley} L.~J.,  {Dopita} M.~A.,  {Sutherland} R.~S.,  {Heisler} C.~A.,
  {Trevena} J.,  2001, \mn@doi [\apj] {10.1086/321545}, \href
  {https://ui.adsabs.harvard.edu/abs/2001ApJ...556..121K} {556, 121}

\bibitem[\protect\citeauthoryear{{Kewley}, {Groves}, {Kauffmann}  \&
  {Heckman}}{{Kewley} et~al.}{2006}]{2006MNRAS.372..961K}
{Kewley} L.~J.,  {Groves} B.,  {Kauffmann} G.,   {Heckman} T.,  2006, \mn@doi
  [\mnras] {10.1111/j.1365-2966.2006.10859.x}, \href
  {https://ui.adsabs.harvard.edu/abs/2006MNRAS.372..961K} {372, 961}

\bibitem[\protect\citeauthoryear{{Law} et~al.,}{{Law}
  et~al.}{2015}]{2015AJ....150...19L}
{Law} D.~R.,  et~al., 2015, \mn@doi [\aj] {10.1088/0004-6256/150/1/19}, \href
  {https://ui.adsabs.harvard.edu/abs/2015AJ....150...19L} {150, 19}

\bibitem[\protect\citeauthoryear{{Law} et~al.,}{{Law}
  et~al.}{2016}]{2016AJ....152...83L}
{Law} D.~R.,  et~al., 2016, \mn@doi [\aj] {10.3847/0004-6256/152/4/83}, \href
  {https://ui.adsabs.harvard.edu/abs/2016AJ....152...83L} {152, 83}

\bibitem[\protect\citeauthoryear{{Lee} et~al.,}{{Lee}
  et~al.}{2015}]{2015ApJ...801...80L}
{Lee} N.,  et~al., 2015, \mn@doi [\apj] {10.1088/0004-637X/801/2/80}, \href
  {https://ui.adsabs.harvard.edu/abs/2015ApJ...801...80L} {801, 80}

\bibitem[\protect\citeauthoryear{{Liang}, {Li}, {Li}, {Yan}, {Mo}, {Zhang},
  {Machuca}  \& {Roman-Lopes}}{{Liang} et~al.}{2020}]{2020ApJ...896..121L}
{Liang} F.-H.,  {Li} C.,  {Li} N.,  {Yan} R.,  {Mo} H.,  {Zhang} W.,  {Machuca}
  C.,   {Roman-Lopes} A.,  2020, \mn@doi [\apj] {10.3847/1538-4357/ab9596},
  \href {https://ui.adsabs.harvard.edu/abs/2020ApJ...896..121L} {896, 121}

\bibitem[\protect\citeauthoryear{{Mannucci} et~al.,}{{Mannucci}
  et~al.}{2022}]{2022NatAs.tmp..187M}
{Mannucci} F.,  et~al., 2022, \mn@doi [Nature Astronomy]
  {10.1038/s41550-022-01761-5}, \href
  {https://ui.adsabs.harvard.edu/abs/2022NatAs.tmp..187M} {}

\bibitem[\protect\citeauthoryear{{Marasco} et~al.,}{{Marasco}
  et~al.}{2020}]{2020A&A...644A..15M}
{Marasco} A.,  et~al., 2020, \mn@doi [\aap] {10.1051/0004-6361/202038889},
  \href {https://ui.adsabs.harvard.edu/abs/2020A&A...644A..15M} {644, A15}

\bibitem[\protect\citeauthoryear{{Marinacci} et~al.,}{{Marinacci}
  et~al.}{2018}]{2018MNRAS.480.5113M}
{Marinacci} F.,  et~al., 2018, \mn@doi [\mnras] {10.1093/mnras/sty2206}, \href
  {https://ui.adsabs.harvard.edu/abs/2018MNRAS.480.5113M} {480, 5113}

\bibitem[\protect\citeauthoryear{{McElroy} et~al.,}{{McElroy}
  et~al.}{2016}]{2016A&A...593L...8M}
{McElroy} R.~E.,  et~al., 2016, \mn@doi [\aap] {10.1051/0004-6361/201629102},
  \href {https://ui.adsabs.harvard.edu/abs/2016A&A...593L...8M} {593, L8}

\bibitem[\protect\citeauthoryear{{Mezcua} \& {Dom{\'\i}nguez
  S{\'a}nchez}}{{Mezcua} \& {Dom{\'\i}nguez
  S{\'a}nchez}}{2020}]{2020ApJ...898L..30M}
{Mezcua} M.,  {Dom{\'\i}nguez S{\'a}nchez} H.,  2020, \mn@doi [\apjl]
  {10.3847/2041-8213/aba199}, \href
  {https://ui.adsabs.harvard.edu/abs/2020ApJ...898L..30M} {898, L30}

\bibitem[\protect\citeauthoryear{{Mezcua}, {Civano}, {Fabbiano}, {Miyaji}  \&
  {Marchesi}}{{Mezcua} et~al.}{2016}]{2016ApJ...817...20M}
{Mezcua} M.,  {Civano} F.,  {Fabbiano} G.,  {Miyaji} T.,   {Marchesi} S.,
  2016, \mn@doi [\apj] {10.3847/0004-637X/817/1/20}, \href
  {https://ui.adsabs.harvard.edu/abs/2016ApJ...817...20M} {817, 20}

\bibitem[\protect\citeauthoryear{{Molina}, {Reines}, {Latimer}, {Baldassare}
  \& {Salehirad}}{{Molina} et~al.}{2021}]{2021ApJ...922..155M}
{Molina} M.,  {Reines} A.~E.,  {Latimer} L.~J.,  {Baldassare} V.,   {Salehirad}
  S.,  2021, \mn@doi [\apj] {10.3847/1538-4357/ac1ffa}, \href
  {https://ui.adsabs.harvard.edu/abs/2021ApJ...922..155M} {922, 155}

\bibitem[\protect\citeauthoryear{{Naiman} et~al.,}{{Naiman}
  et~al.}{2018}]{2018MNRAS.477.1206N}
{Naiman} J.~P.,  et~al., 2018, \mn@doi [\mnras] {10.1093/mnras/sty618}, \href
  {https://ui.adsabs.harvard.edu/abs/2018MNRAS.477.1206N} {477, 1206}

\bibitem[\protect\citeauthoryear{{Nakajima} \& {Maiolino}}{{Nakajima} \&
  {Maiolino}}{2022}]{2022MNRAS.513.5134N}
{Nakajima} K.,  {Maiolino} R.,  2022, \mn@doi [\mnras]
  {10.1093/mnras/stac1242}, \href
  {https://ui.adsabs.harvard.edu/abs/2022MNRAS.513.5134N} {513, 5134}

\bibitem[\protect\citeauthoryear{{Negus}, {Comerford}, {M{\"u}ller
  S{\'a}nchez}, {Barrera-Ballesteros}, {Drory}, {Rembold}  \& {Riffel}}{{Negus}
  et~al.}{2021}]{2021ApJ...920...62N}
{Negus} J.,  {Comerford} J.~M.,  {M{\"u}ller S{\'a}nchez} F.,
  {Barrera-Ballesteros} J.~K.,  {Drory} N.,  {Rembold} S.~B.,   {Riffel} R.~A.,
   2021, \mn@doi [\apj] {10.3847/1538-4357/ac1343}, \href
  {https://ui.adsabs.harvard.edu/abs/2021ApJ...920...62N} {920, 62}

\bibitem[\protect\citeauthoryear{{Negus}, {Comerford}, {Muller Sanchez},
  {Revalski}, {Riffel}, {Bundy}, {Nevin}  \& {Rembold}}{{Negus}
  et~al.}{2023}]{2023arXiv230113322N}
{Negus} J.,  {Comerford} J.~M.,  {Muller Sanchez} F.,  {Revalski} M.,  {Riffel}
  R.~A.,  {Bundy} K.,  {Nevin} R.,   {Rembold} S.~B.,  2023, \mn@doi [arXiv
  e-prints] {10.48550/arXiv.2301.13322}, \href
  {https://ui.adsabs.harvard.edu/abs/2023arXiv230113322N} {p. arXiv:2301.13322}

\bibitem[\protect\citeauthoryear{{Nelson} et~al.,}{{Nelson}
  et~al.}{2018}]{2018MNRAS.475..624N}
{Nelson} D.,  et~al., 2018, \mn@doi [\mnras] {10.1093/mnras/stx3040}, \href
  {https://ui.adsabs.harvard.edu/abs/2018MNRAS.475..624N} {475, 624}

\bibitem[\protect\citeauthoryear{{Noeske} et~al.,}{{Noeske}
  et~al.}{2007}]{2007ApJ...660L..43N}
{Noeske} K.~G.,  et~al., 2007, \mn@doi [\apjl] {10.1086/517926}, \href
  {https://ui.adsabs.harvard.edu/abs/2007ApJ...660L..43N} {660, L43}

\bibitem[\protect\citeauthoryear{{Pillepich} et~al.,}{{Pillepich}
  et~al.}{2018}]{2018MNRAS.475..648P}
{Pillepich} A.,  et~al., 2018, \mn@doi [\mnras] {10.1093/mnras/stx3112}, \href
  {https://ui.adsabs.harvard.edu/abs/2018MNRAS.475..648P} {475, 648}

\bibitem[\protect\citeauthoryear{{Piotrowska}, {Bluck}, {Maiolino}  \&
  {Peng}}{{Piotrowska} et~al.}{2022}]{2022MNRAS.512.1052P}
{Piotrowska} J.~M.,  {Bluck} A. F.~L.,  {Maiolino} R.,   {Peng} Y.,  2022,
  \mn@doi [\mnras] {10.1093/mnras/stab3673}, \href
  {https://ui.adsabs.harvard.edu/abs/2022MNRAS.512.1052P} {512, 1052}

\bibitem[\protect\citeauthoryear{{Reefe}, {Satyapal}, {Sexton}, {Doan},
  {Secrest}  \& {Cann}}{{Reefe} et~al.}{2022}]{2022ApJ...936..140R}
{Reefe} M.,  {Satyapal} S.,  {Sexton} R.~O.,  {Doan} S.~M.,  {Secrest} N.~J.,
  {Cann} J.~M.,  2022, \mn@doi [\apj] {10.3847/1538-4357/ac8981}, \href
  {https://ui.adsabs.harvard.edu/abs/2022ApJ...936..140R} {936, 140}

\bibitem[\protect\citeauthoryear{{R{\"o}ck}, {Vazdekis}, {Ricciardelli},
  {Peletier}, {Knapen}  \& {Falc{\'o}n-Barroso}}{{R{\"o}ck}
  et~al.}{2016}]{2016A&A...589A..73R}
{R{\"o}ck} B.,  {Vazdekis} A.,  {Ricciardelli} E.,  {Peletier} R.~F.,  {Knapen}
  J.~H.,   {Falc{\'o}n-Barroso} J.,  2016, \mn@doi [\aap]
  {10.1051/0004-6361/201527570}, \href
  {https://ui.adsabs.harvard.edu/abs/2016A&A...589A..73R} {589, A73}

\bibitem[\protect\citeauthoryear{{Rogers}, {Skillman}, {Pogge}, {Berg},
  {Moustakas}, {Croxall}  \& {Sun}}{{Rogers}
  et~al.}{2021}]{2021ApJ...915...21R}
{Rogers} N. S.~J.,  {Skillman} E.~D.,  {Pogge} R.~W.,  {Berg} D.~A.,
  {Moustakas} J.,  {Croxall} K.~V.,   {Sun} J.,  2021, \mn@doi [\apj]
  {10.3847/1538-4357/abf8b9}, \href
  {https://ui.adsabs.harvard.edu/abs/2021ApJ...915...21R} {915, 21}

\bibitem[\protect\citeauthoryear{{S{\'a}nchez} et~al.,}{{S{\'a}nchez}
  et~al.}{2016}]{2016RMxAA..52..171S}
{S{\'a}nchez} S.~F.,  et~al., 2016, \rmxaa, \href
  {https://ui.adsabs.harvard.edu/abs/2016RMxAA..52..171S} {52, 171}

\bibitem[\protect\citeauthoryear{{Schaerer}, {Fragos}  \& {Izotov}}{{Schaerer}
  et~al.}{2019}]{2019A&A...622L..10S}
{Schaerer} D.,  {Fragos} T.,   {Izotov} Y.~I.,  2019, \mn@doi [\aap]
  {10.1051/0004-6361/201935005}, \href
  {https://ui.adsabs.harvard.edu/abs/2019A&A...622L..10S} {622, L10}

\bibitem[\protect\citeauthoryear{{Schawinski}, {Thomas}, {Sarzi}, {Maraston},
  {Kaviraj}, {Joo}, {Yi}  \& {Silk}}{{Schawinski}
  et~al.}{2007}]{2007MNRAS.382.1415S}
{Schawinski} K.,  {Thomas} D.,  {Sarzi} M.,  {Maraston} C.,  {Kaviraj} S.,
  {Joo} S.-J.,  {Yi} S.~K.,   {Silk} J.,  2007, \mn@doi [\mnras]
  {10.1111/j.1365-2966.2007.12487.x}, \href
  {https://ui.adsabs.harvard.edu/abs/2007MNRAS.382.1415S} {382, 1415}

\bibitem[\protect\citeauthoryear{{Schawinski}, {Virani}, {Simmons}, {Urry},
  {Treister}, {Kaviraj}  \& {Kushkuley}}{{Schawinski}
  et~al.}{2009}]{2009ApJ...692L..19S}
{Schawinski} K.,  {Virani} S.,  {Simmons} B.,  {Urry} C.~M.,  {Treister} E.,
  {Kaviraj} S.,   {Kushkuley} B.,  2009, \mn@doi [\apjl]
  {10.1088/0004-637X/692/1/L19}, \href
  {https://ui.adsabs.harvard.edu/abs/2009ApJ...692L..19S} {692, L19}

\bibitem[\protect\citeauthoryear{{Schawinski} et~al.,}{{Schawinski}
  et~al.}{2010}]{2010ApJ...711..284S}
{Schawinski} K.,  et~al., 2010, \mn@doi [\apj] {10.1088/0004-637X/711/1/284},
  \href {https://ui.adsabs.harvard.edu/abs/2010ApJ...711..284S} {711, 284}

\bibitem[\protect\citeauthoryear{{Schaye} et~al.,}{{Schaye}
  et~al.}{2015}]{2015MNRAS.446..521S}
{Schaye} J.,  et~al., 2015, \mn@doi [\mnras] {10.1093/mnras/stu2058}, \href
  {https://ui.adsabs.harvard.edu/abs/2015MNRAS.446..521S} {446, 521}

\bibitem[\protect\citeauthoryear{{Schneider} et~al.,}{{Schneider}
  et~al.}{2010}]{2010AJ....139.2360S}
{Schneider} D.~P.,  et~al., 2010, \mn@doi [\aj] {10.1088/0004-6256/139/6/2360},
  \href {https://ui.adsabs.harvard.edu/abs/2010AJ....139.2360S} {139, 2360}

\bibitem[\protect\citeauthoryear{{Schutte} \& {Reines}}{{Schutte} \&
  {Reines}}{2022}]{2022Natur.601..329S}
{Schutte} Z.,  {Reines} A.~E.,  2022, \mn@doi [\nat]
  {10.1038/s41586-021-04215-6}, \href
  {https://ui.adsabs.harvard.edu/abs/2022Natur.601..329S} {601, 329}

\bibitem[\protect\citeauthoryear{{Shapovalova}, {Popovi{\'c}}, {Burenkov},
  {Chavushyan}, {Ili{\'c}}, {Kova{\v{c}}evi{\'c}}, {Bochkarev}  \&
  {Le{\'o}n-Tavares}}{{Shapovalova} et~al.}{2010}]{2010A&A...509A.106S}
{Shapovalova} A.~I.,  {Popovi{\'c}} L.~{\v{C}}.,  {Burenkov} A.~N.,
  {Chavushyan} V.~H.,  {Ili{\'c}} D.,  {Kova{\v{c}}evi{\'c}} A.,  {Bochkarev}
  N.~G.,   {Le{\'o}n-Tavares} J.,  2010, \mn@doi [\aap]
  {10.1051/0004-6361/200912311}, \href
  {https://ui.adsabs.harvard.edu/abs/2010A&A...509A.106S} {509, A106}

\bibitem[\protect\citeauthoryear{{Shirazi} \& {Brinchmann}}{{Shirazi} \&
  {Brinchmann}}{2012}]{2012MNRAS.421.1043S}
{Shirazi} M.,  {Brinchmann} J.,  2012, \mn@doi [\mnras]
  {10.1111/j.1365-2966.2012.20439.x}, \href
  {https://ui.adsabs.harvard.edu/abs/2012MNRAS.421.1043S} {421, 1043}

\bibitem[\protect\citeauthoryear{{Springel} et~al.,}{{Springel}
  et~al.}{2018}]{2018MNRAS.475..676S}
{Springel} V.,  et~al., 2018, \mn@doi [\mnras] {10.1093/mnras/stx3304}, \href
  {https://ui.adsabs.harvard.edu/abs/2018MNRAS.475..676S} {475, 676}

\bibitem[\protect\citeauthoryear{{Stern} et~al.,}{{Stern}
  et~al.}{2005}]{2005ApJ...631..163S}
{Stern} D.,  et~al., 2005, \mn@doi [\apj] {10.1086/432523}, \href
  {https://ui.adsabs.harvard.edu/abs/2005ApJ...631..163S} {631, 163}

\bibitem[\protect\citeauthoryear{{Stern} et~al.,}{{Stern}
  et~al.}{2012}]{2012ApJ...753...30S}
{Stern} D.,  et~al., 2012, \mn@doi [\apj] {10.1088/0004-637X/753/1/30}, \href
  {https://ui.adsabs.harvard.edu/abs/2012ApJ...753...30S} {753, 30}

\bibitem[\protect\citeauthoryear{{Thuan} \& {Izotov}}{{Thuan} \&
  {Izotov}}{2005}]{2005ApJS..161..240T}
{Thuan} T.~X.,  {Izotov} Y.~I.,  2005, \mn@doi [\apjs] {10.1086/491657}, \href
  {https://ui.adsabs.harvard.edu/abs/2005ApJS..161..240T} {161, 240}

\bibitem[\protect\citeauthoryear{{Tozzi} et~al.,}{{Tozzi}
  et~al.}{2021}]{2021A&A...648A..99T}
{Tozzi} G.,  et~al., 2021, \mn@doi [\aap] {10.1051/0004-6361/202040190}, \href
  {https://ui.adsabs.harvard.edu/abs/2021A&A...648A..99T} {648, A99}

\bibitem[\protect\citeauthoryear{{Trussler}, {Maiolino}, {Maraston}, {Peng},
  {Thomas}, {Goddard}  \& {Lian}}{{Trussler}
  et~al.}{2020}]{2020MNRAS.491.5406T}
{Trussler} J.,  {Maiolino} R.,  {Maraston} C.,  {Peng} Y.,  {Thomas} D.,
  {Goddard} D.,   {Lian} J.,  2020, \mn@doi [\mnras] {10.1093/mnras/stz3286},
  \href {https://ui.adsabs.harvard.edu/abs/2020MNRAS.491.5406T} {491, 5406}

\bibitem[\protect\citeauthoryear{{Umeda}, {Ouchi}, {Nakajima}, {Isobe},
  {Aoyama}, {Harikane}, {Ono}  \& {Matsumoto}}{{Umeda}
  et~al.}{2022}]{2022ApJ...930...37U}
{Umeda} H.,  {Ouchi} M.,  {Nakajima} K.,  {Isobe} Y.,  {Aoyama} S.,  {Harikane}
  Y.,  {Ono} Y.,   {Matsumoto} A.,  2022, \mn@doi [\apj]
  {10.3847/1538-4357/ac602d}, \href
  {https://ui.adsabs.harvard.edu/abs/2022ApJ...930...37U} {930, 37}

\bibitem[\protect\citeauthoryear{{Wadadekar}}{{Wadadekar}}{2004}]{2004A&A...416...35W}
{Wadadekar} Y.,  2004, \mn@doi [\aap] {10.1051/0004-6361:20034244}, \href
  {https://ui.adsabs.harvard.edu/abs/2004A&A...416...35W} {416, 35}

\bibitem[\protect\citeauthoryear{{Wake} et~al.,}{{Wake}
  et~al.}{2017}]{2017AJ....154...86W}
{Wake} D.~A.,  et~al., 2017, \mn@doi [\aj] {10.3847/1538-3881/aa7ecc}, \href
  {https://ui.adsabs.harvard.edu/abs/2017AJ....154...86W} {154, 86}

\bibitem[\protect\citeauthoryear{{Ward}, {Harrison}, {Costa}  \&
  {Mainieri}}{{Ward} et~al.}{2022}]{2022MNRAS.514.2936W}
{Ward} S.~R.,  {Harrison} C.~M.,  {Costa} T.,   {Mainieri} V.,  2022, \mn@doi
  [\mnras] {10.1093/mnras/stac1219}, \href
  {https://ui.adsabs.harvard.edu/abs/2022MNRAS.514.2936W} {514, 2936}

\bibitem[\protect\citeauthoryear{{Westfall} et~al.,}{{Westfall}
  et~al.}{2019}]{2019AJ....158..231W}
{Westfall} K.~B.,  et~al., 2019, \mn@doi [\aj] {10.3847/1538-3881/ab44a2},
  \href {https://ui.adsabs.harvard.edu/abs/2019AJ....158..231W} {158, 231}

\bibitem[\protect\citeauthoryear{{Wylezalek} et~al.,}{{Wylezalek}
  et~al.}{2017}]{2017MNRAS.467.2612W}
{Wylezalek} D.,  et~al., 2017, \mn@doi [\mnras] {10.1093/mnras/stx246}, \href
  {https://ui.adsabs.harvard.edu/abs/2017MNRAS.467.2612W} {467, 2612}

\bibitem[\protect\citeauthoryear{{Yan} et~al.,}{{Yan}
  et~al.}{2016}]{2016AJ....152..197Y}
{Yan} R.,  et~al., 2016, \mn@doi [\aj] {10.3847/0004-6256/152/6/197}, \href
  {https://ui.adsabs.harvard.edu/abs/2016AJ....152..197Y} {152, 197}

\bibitem[\protect\citeauthoryear{{York} et~al.,}{{York}
  et~al.}{2000}]{2000AJ....120.1579Y}
{York} D.~G.,  et~al., 2000, \mn@doi [\aj] {10.1086/301513}, \href
  {https://ui.adsabs.harvard.edu/abs/2000AJ....120.1579Y} {120, 1579}

\makeatother
\end{thebibliography}



\appendix
\section{Spatially resolved diagnostic maps of the \ion{He}{ii}-only AGN candidates}
\label{appx1}

\begin{figure*}
	\includegraphics[width=1.9\columnwidth]{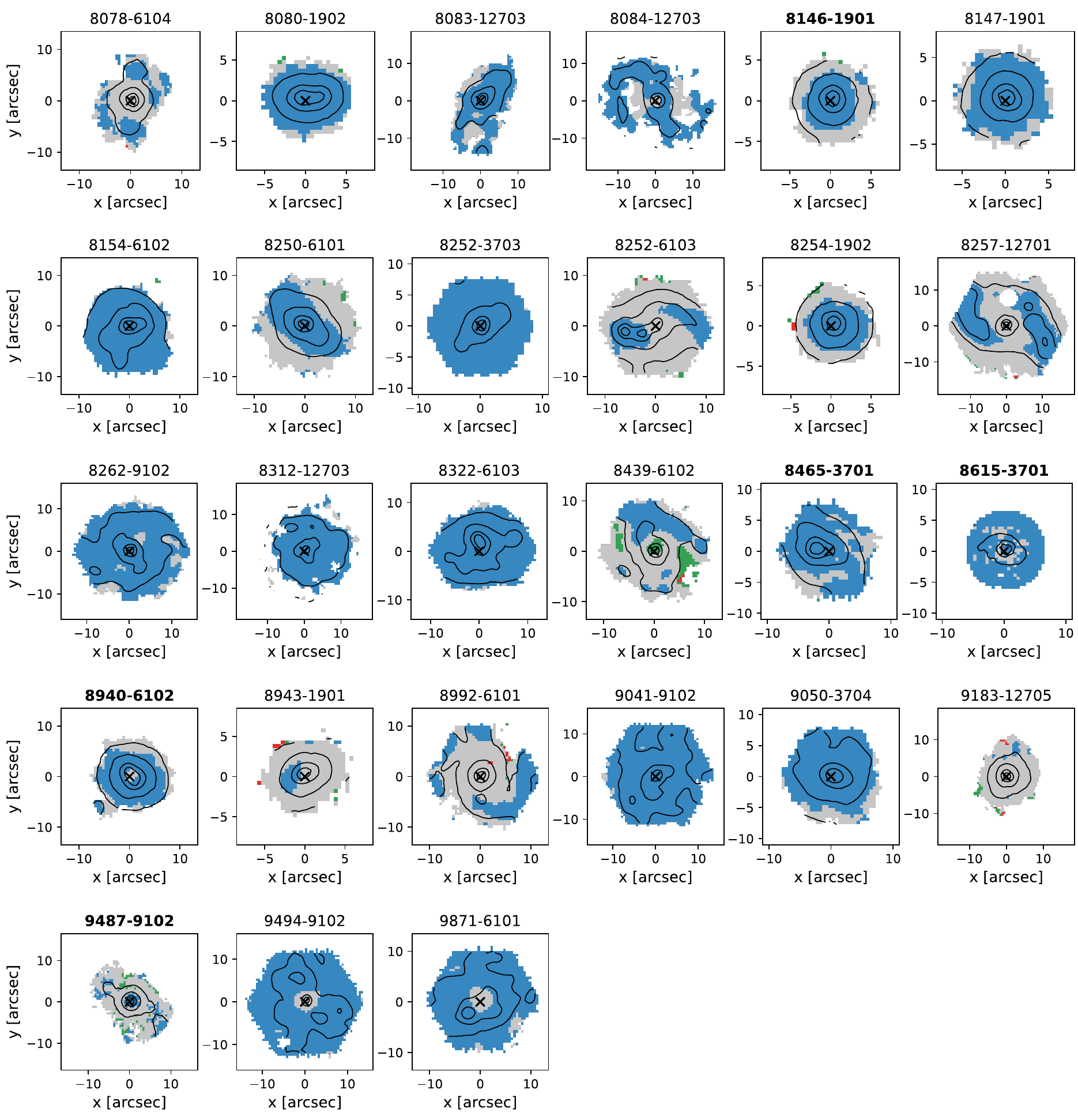}
    \caption{Spatially resolved BPT maps of the 27 \ion{He}{ii}-only AGN galaxies resulting from our selection (Sect. \ref{subsec:AGN_galaxies}). Distinct colours identify regions ionised by different mechanisms \citep{2001ApJ...556..121K, 2003MNRAS.346.1055K, 2007MNRAS.382.1415S}: star formation (blue), LINER (green), composite processes (grey), and AGN (red, globally absent). Black lines are contour levels of H$\alpha$ line flux corresponding to 1\%, 10\%, 50\% and 80\% values of the emission peak, and the black cross marks the galaxy centre. Coloured spaxels are those meeting the S/N threshold (Sect. \ref{subsec:eldiag}), and not masked by the MaNGA DAP. Bold labels indicate galaxies whose MaNGA datacubes needed some spectral remodelling (see Table \ref{tab1}).}
    \label{fig_a1}
\end{figure*}

\begin{figure*}
	\includegraphics[width=1.9\columnwidth]{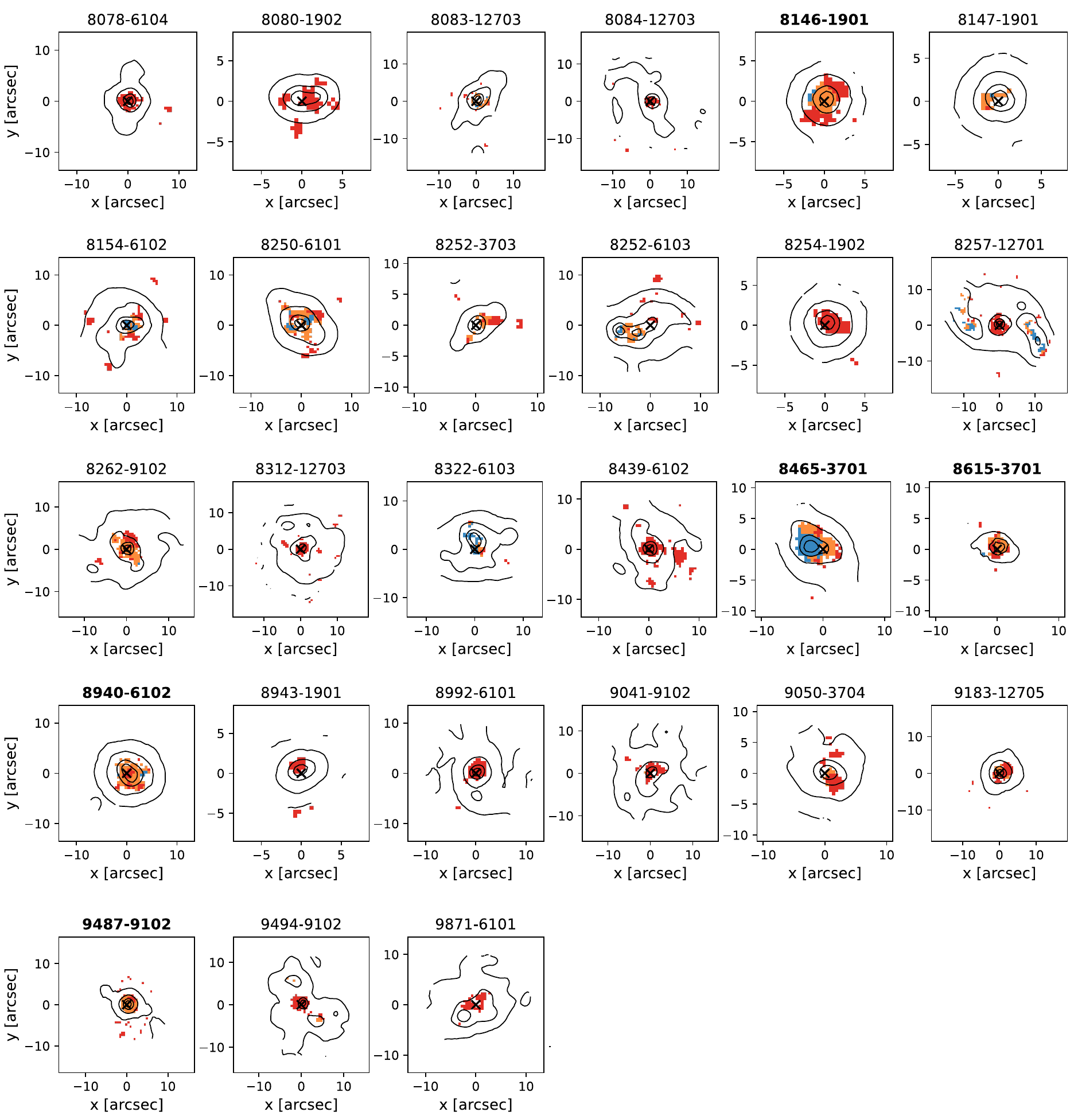}
    \caption{Spatially resolved \ion{He}{ii} maps of the 27 \ion{He}{ii}-only AGN galaxies resulting from our selection (Sect. \ref{subsec:AGN_galaxies}). Distinct colours identify regions ionised by different mechanisms \citep{2012MNRAS.421.1043S}: star formation (blue), strong AGN (red, above the 50\% curve; see Fig. \ref{fig1} and related text), and weak AGN (orange, above the 10\% curve). Black lines, black cross and bold labels have the same meaning as in Fig. \ref{fig_a1}. Coloured spaxels fulfil the S/N threshold (Sect. \ref{subsec:eldiag}), and are not masked by the MaNGA DAP.}
    \label{fig_a2}
\end{figure*}

In Figs. \ref{fig_a1} and \ref{fig_a2}, we show spatially resolved maps of every selected \ion{He}{ii}-only AGN galaxy, displaying the spaxel classification according to BPT (\citealt{1981PASP...93....5B}; Fig. \ref{fig_a1}) and \ion{He}{ii} (\citealt{2012MNRAS.421.1043S}; Fig. \ref{fig_a2}) diagrams, respectively. Bold labels indicate galaxies whose MaNGA datacubes needed some spectral remodelling (see Table \ref{tab1}). In BPT maps (Fig. \ref{fig_a1}), we use distinct colours to distinguish different ionisation mechanisms \citep{2001ApJ...556..121K, 2003MNRAS.346.1055K, 2007MNRAS.382.1415S}: star formation (blue), LINER (green), composite mechanisms (grey), and AGN (red, globally absent). With the exception of very few isolated red spaxels in galaxy outskirts (totally negligible), the BPT diagnostic detects no AGN-like emission in these galaxies, in agreement with our BPT-based classification of these objects as non-active galaxies.

In a similar way, we use different colours to indicate regions ionised by distinct mechanisms according to the \ion{He}{ii} diagram (Fig. \ref{fig_a2}): star formation (blue) and AGN, for which we distinguish between a stronger (red, above the 50\% curve; see Fig. \ref{fig1} an related text) and weaker (orange, above the 10\% curve) AGN contribution to the \ion{He}{ii} \citep{2012MNRAS.421.1043S}. In both BPT and \ion{He}{ii} maps, solid black lines are contour levels of H$\alpha$ line flux corresponding to 1\%, 10\%, 50\% and 80\% values of the respective emission peak, and a black cross marks the galaxy centre. Coloured spaxels are those meeting the S/N threshold, as defined at the beginning of Sect. \ref{subsec:eldiag}. In addition, we apply MaNGA DAP masks associated with the flux each emission line, which explains why in a few maps of Fig. \ref{fig_a2} the number of red (strong AGN) spaxels is smaller than the 20-spaxel threshold adopted for our preliminary selection of AGN galaxies (Sect. \ref{subsec:AGN_galaxies}). Yet, we retain such galaxies in the \ion{He}{ii} AGN sample since they however show centrally-located, \ion{He}{ii}-classified AGN emission.


\bsp	
\label{lastpage}
\end{document}